# Detection of large-scale noisy multi-periodic patterns with discrete double Fourier transform

V.R. Chechetkin[a, b]* and V.V. Lobzin[c]

[a]*Theoretical Department of Division for Perspective Investigations, Troitsk Institute of Innovation and Thermonuclear Investigations (TRINITI), Moscow, Troitsk District 108840, Russia*

[b]*Engelhardt Institute of Molecular Biology of Russian Academy of Sciences, Vavilov str., 32, Moscow 119334, Russia*

[c]*School of Physics, University of Sydney, Sydney, NSW 2006, Australia*

___________________________

*Corresponding author.

*E-mail addresses:* chechet@eimb.ru; vladimir_chechet@mail.ru (V.R. Chechetkin); vasili.lobzin@sydney.edu.au (V.V. Lobzin).

In many processes, the variations in underlying characteristics can be approximated by noisy multi-periodic patterns. If large-scale patterns are superimposed by a noise with long-range correlations, the detection of multi-periodic patterns becomes especially challenging. To solve this problem, we developed a discrete double Fourier transform (DDFT). DDFT is based on the equidistance property of harmonics generated by multi-periodic patterns in the discrete Fourier transform (DFT) spectra. As the large-scale patterns generate long enough equidistant series, they can be detected by the iteration of the primary DFT. DDFT is defined as Fourier transform of intensity spectral harmonics or of their functions. It comprises widely used cepstrum transform as a particular case. We present also the relevant analytical criteria for the assessment of statistical significance of peak harmonics in DDFT spectra in the presence of noise. DDFT technique was tested by extensive numerical simulations. The practical applications of DDFT technique are illustrated by the analysis of variations in solar wind speed related to solar rotation and by the study of large-scale multi-periodic patterns in DNA sequences. The latter application can be considered as generic example for the general spectral analysis of symbolic sequences. The results are compared with those obtained by the cepstrum transform. The mutual combination of DFT and DDFT provides an efficient technique to search for noisy large-scale multi-periodic patterns.

# 1. Introduction

Noisy periodic and multi-periodic (composed of superimposed multiple periods) patterns are ubiquitous in many engineering, physical, geophysical, astronomical, medical, genetic, economic, and financial problems (see, e.g., [1–11] and further references therein). Fourier and wavelet transforms provide currently the most common and efficient tools for detection of multi-periodic patterns in noisy data sets [12–21]. The robustness of the both methods is approximately the same, though the sensitivity and the precision of period resolution in Fourier transform is higher. Fourier transform is also more convenient for the study of multi-periodic rather than purely periodic patterns and for the study of superimposed patterns with periods of comparable lengths. The resonant correspondence between a period of pattern and that of sines/cosines in the Fourier basis produces the peak harmonics in the related Fourier spectrum. The search for noisy multi-periodic patterns is much more difficult and needs the sophisticated processing schemes and statistical methods. The statistical assessment of periodic features detected via wavelet transform needs the extensive numerical simulations, whereas the peak harmonics in Fourier spectra can be assessed via relevant analytical criteria that provides the additional advantages. The detection of large-scale noisy multi-periodic patterns with Fourier transform remains yet especially challenging because the multi-periodic contributions into the harmonics in the range of low spectral numbers are commonly strongly superimposed on the contributions from unrelated sources. The latter are mainly due to: (i) stochastic $1/f^{\alpha}$ noise; (ii) large-scale non-periodic variations in the values of data over set; and (iii) long-range correlations in underlying data. All these sources generate high harmonics in the range of low spectral numbers with heights decreasing with the increase of spectral number and hamper the detection of large-scale periodicities. The other class of problems is related to large-scale modulations of short-periodic patterns, e.g., the pulses in cardiograms or encephalograms are superimposed with modulations related to one-day activity (circadian rhythms). The numerous examples of modulations in radiophysics, solar and plasma physics, DNA sequences, etc. may easily be added to this list. The method presented below comprises such a class of problems as well.

Besides discrete Fourier transform (DFT) and wavelet technique, the family of applied transforms includes also the Walsh-Hadamard [22, 23] and Ramanujan periodicity transforms (RPT) [24–28]. The Walsh-Hadamard transform is limited to the total number of points in the data set of the form $N = 2^k$ and has approximately the same applicability as the fast Fourier transform. RPT is based on the expansion of data set over the Ramanujan sums. When choosing a particular transform to detect noisy multi-periodic patterns, the user relies often on the expected unique correspondence peak-period in the resulting spectra. It is now clearly understood that this is not a generic case and any high spectral peak needs an additional verification. The further progress in search for noisy multi-periodic patterns is based on the decomposition of spectrum over signal spectral subspaces [26, 29–31]. Except purely sin/cos periodic patterns, all other periodic patterns of different shape generate a series of equidistant peak harmonics in DFT spectrum (see, e.g., [8] and Section 2.2 below). The series of equidistant harmonics form the natural signal subspaces in DFT spectra. Unlike RPT, where the periods should be strictly integer, the related periodicities in DFT can be generally non-integer. The latter feature may be important, because in many natural problems the distribution of observable variations may approximate the actual non-integer periodicities. In particular, the helix pitch in the double-stranded B-form DNA varies within 10.2–10.8 base pairs, which may be concordant with underlying distribution of nucleotides along the genome and may be related with various genetic regulation mechanisms (see, e.g., [32, 33] and references therein). The resulting signal subspaces in RPT strongly depend on the divisors of $N$ [26]. The Ramanujan sums used as basis functions in RPT are non-orthogonal in $N$-dimensional data space. The computation of related expansion coefficients in RPT needs the inversion of $N \times N$ matrices. The complexity of computations grows rapidly with the increase of $N$. As the basis functions in DFT are orthogonal in $N$-dimensional space, the corresponding matrices in DFT are diagonal and can be easily inverted. We will show that the projection of DFT spectrum onto signal subspaces related to large-scale multi-periodic patterns can be efficiently performed via discrete double Fourier transform (DDFT). For large-scale periodic patterns the related series of equidistant harmonics are long enough to be detected by the iteration of Fourier transform (or DDFT). In this paper we present general formulation and further extension of DDFT technique developed previously [34]. We also present some analytical criteria related to the applications of DDFT as well as the results of extensive numerical simulations aimed at the assessment of DDFT robustness to noise.

## 2. Theory and methods

### 2.1. Discrete Fourier transform (DFT)

In this section we summarize briefly main results concerning DFT needed for the subsequent analysis (see, e.g., [15, 18, 19]). Our consideration is restricted to real-valued variables $x_n$. They may correspond either to the discretized version of continuous variables or to two-valued Boolean variables taking the values 1 and 0. The latter version can be applied to the Fourier analysis of symbolic sequences (see, e.g., [35, 36] and Section 4.2 below). The direct and inverse DFT for the set of $N$ sampling points are defined as

$$X_k = N^{-1/2} \sum_{n=0}^{N-1} x_n e^{-i2\pi kn/N}, \quad k = 0, 1, ..., N-1 \tag{1}$$

$$x_n = N^{-1/2} \sum_{k=0}^{N-1} X_k e^{i2\pi kn/N}, \quad n = 0, 1, ..., N-1 \tag{2}$$

Taking into account the equality

$$\sum_{n=0}^{N-1} e^{-i2\pi nk/N} = \begin{cases} 0, \text{ if } k = 1, 2, ..., N-1 \\ N, \text{ if } k = 0 \end{cases} \tag{3}$$

the harmonics with $k \neq 0$ can also be redefined as

$$X_k = N^{-1/2} \sum_{n=0}^{N-1} \Delta x_n e^{-i2\pi kn/N}, \quad k = 1, 2, ..., N-1 \tag{4}$$

where

$$\Delta x_n = x_n - \bar{x}; \quad \bar{x} = N^{-1} \sum_{n=0}^{N-1} x_n \tag{5}$$

The spectral intensity is given by

$$F_k = X_k X_k^* \tag{6}$$

where the asterisk denotes the complex conjugation. The harmonic (1) with $k = 0$ depends only on the mean value of $x_n$ and on the total number of sampling points ($X_0 = \bar{x} N^{1/2}$) and is insensitive to periodic variations in underlying data, whereas the other components reflect the variations around mean value (see Eq. (4)). As detection of multi-periodic patterns is related only to the variations, below we will restrict ourselves to harmonics with $k \neq 0$.

For real-valued $x_n$ the harmonics $X_k$ with the different spectral numbers are related as

$$X_k = X_{N-k}^* \tag{7}$$

This yields the symmetry relationship for the intensity harmonics,

$$F_k = F_{N-k} \tag{8}$$

In this case Fourier spectrum can be restricted from $k = 1$ to

$$K = [N/2] \tag{9}$$

where the brackets denote the integer part of the quotient. The mean spectral intensity can be expressed as

$$\overline{F} = \frac{1}{N-1}\sum_{k=1}^{N-1} X_k X_k^* = \frac{1}{N-1}\sum_{n=0}^{N-1}(\Delta x_n)^2 = \sigma^2 \tag{10}$$

It is convenient to normalize the harmonics for spectral intensity on the mean value

$$f_k = F_k / \overline{F} \tag{11}$$

The normalized harmonics $f_k$ with $k \neq 0$ remain invariant under affine transforms of variables $\{x_n\}$ in Eq. (1), $x_n \to a x_n + b$. The periods of patterns can be assessed as

$$p = N/k \tag{12}$$

## 2.2. Discrete double Fourier transform (DDFT)
### 2.2.1. Motivation

Consider now the data set composed of $N_P$ periodic patterns of period $P$, such that

$$PN_P = N \tag{13}$$

In the periodic patterns of period $P$ the values of $x_n$ are related as $x_0 = x_P$, $x_1 = x_{P+1}$, ..., $x_{P-1} = x_{2P-1}$, etc. Therefore, the corresponding Fourier harmonics can be presented as

$$X_k = N^{-1/2}\sum_{n=0}^{N-1} x_n e^{-i2\pi kn/N} = \left(N_P^{-1/2}\sum_{n=0}^{N_P-1} e^{-i2\pi kPn/N}\right)\left(P^{-1/2}\sum_{n=0}^{P-1} x_n e^{-i2\pi kn/N}\right) \equiv \chi_{k;P} X_{k;P} \tag{14}$$

and the spectral intensity is given by

$$F_k = \Phi_{k;P} F_{k;P} \tag{15}$$

where

$$\Phi_{k;P} = \chi_{k;P}\chi_{k;P}^*;\ F_{k;P} = X_{k;P} X_{k;P}^* \tag{16}$$

The factor $\Phi_{k;P}$ is equal to

$$\Phi_{k;P} = \begin{cases} N_P, & \text{if } k = N/P, 2N/P, ..., (P-1)N/P \\ 0, & \text{otherwise} \end{cases} \tag{17}$$

This means that all periodic patterns except purely sin/cos-like patterns generate a series of equidistant peak harmonics. The averaging of non-zero harmonics (15) over the peaks at $k = N/P, 2N/P, ..., (P-1)N/P$ yields the mean

$$\overline{F_k} = N_P \sigma_P^2 \tag{18}$$

where the variance $\sigma_P^2$ is calculated over repeating pattern similarly to Eq. (10). Due to the symmetry property (8) the series of equidistant peak harmonics can be restricted by $k_{\max} N/P \leq K$. The extreme left harmonic in the equidistant series will be called a generating (or fundamental) harmonic and the related spectral number will be denoted as $k_g$. The relevant spectral number is directly related to the total number of periodic patterns in the data set, $N_P = k_g$. The property of equidistance can be considered as a benchmark of true periodicity, which is absent for $1/f^\alpha$ noise and other non-periodic sources. The series of equidistant harmonics in DFT spectrum for the strict repeats have simple meaning from the point of view of linear algebra and information theory. DFT spectrum for the strict repeats of a period $P$ should code for the values of variables in $P$ positions of repeat (these correspond to $P$ peaks in Fourier spectrum) and for the total number of repeats in the set (which corresponds to the distance between consecutive equidistant harmonics). Therefore, the series of equidistant harmonics in DFT spectrum provide the natural bases for the signal subspaces related with different periodicities. Such equidistant series will persist in the presence of noise as well. If $k_{\max} \gg 1$ (that is fulfilled for the large-scale patterns), the noisy equidistant series can be detected by the iteration of Fourier transform or DDFT.

*2.2.2. Main definitions and formulae for DDFT*

**Definition:** Discrete double Fourier transform is defined as discrete Fourier transform of normalized intensity harmonics $f_k$ (Eq. (11)) (or, generally, of functions $\Phi(f_k)$) in the spectrum without harmonics with the spectral numbers $k = 0$ and 1.

We start with this simplest definition and postpone generalizations of DDFT to Section 2.2.3. DDFT can be performed for the normalized intensity harmonics (11) along the lines of definitions for DFT,

$$X_{k'; II} = (K-1)^{-1/2} \sum_{k=0}^{K-2} f_{k+2} e^{-i2\pi kk'/(K-1)}, \quad k' = 0, 1, ..., K-2 \tag{19}$$

$$F_{k'; II} = X_{k'; II} X_{k'; II}^* \tag{20}$$

As the DFT harmonic with $k=1$ does not generate an equidistant series, it is discarded from DDFT. The symmetry relationship for the intensity harmonics (20) similar to that in Eq. (8) will be valid as well, that allows to restrict the right boundary of DDFT spectrum at

$$K'_{II} = [(K-1)/2] \tag{21}$$

The related normalized intensity harmonics are defined as

$$f_{k';II} = F_{k';II} / \overline{F_{II}} \tag{22}$$

$$\overline{F_{II}} = \frac{1}{K'_{II}} \sum_{k'=1}^{K'_{II}} F_{k';II} \tag{23}$$

Generally, the equidistant series started at $k_g$ in DFT spectrum produces the corresponding equidistant peaks in DDFT spectrum at

$$\{k'_{II}\}_{eq} = k'_{II,g}, 2k'_{II,g}, ... \leq K'_{II}; k'_{II,g} \approx (K-1)/k_g \tag{24}$$

The total number of periodic patterns in the primary data set assessed by the equidistant peaks in DDFT spectrum is given by

$$N'_{p;II} = (K-1)/k'_{II,g} \tag{25}$$

whereas the periods assessed via DDFT are

$$p'_{II} = N/N'_{p;II} \tag{26}$$

Fig. 1 illustrates the related DFT and DDFT spectra for some typical periodic patterns. Only pure sin/cos patterns generate a unique peak in DFT spectrum (Fig. 1A). All other patterns generate a series of equidistant peaks in DFT spectra and the related equidistant peaks in DDFT spectra (Figs. 1B–1D). If the periodic patterns approximately resemble the sin/cos-like patterns, the heights of equidistant peaks in DFT spectra decrease with the increase in spectral numbers (Figs. 1B and 1C). The harmonic at $k_g$ appears to be the highest for such patterns and can be used for the assessment of the total number of periodic patterns in the data set and the period by Eq. (12). For patterns of complicated shape (mimicked by repeating patterns with Boolean or Gaussian variables) the highest peak in DFT spectrum may not be associated with the number of periodic patterns and relevant periods (Fig. 1D).

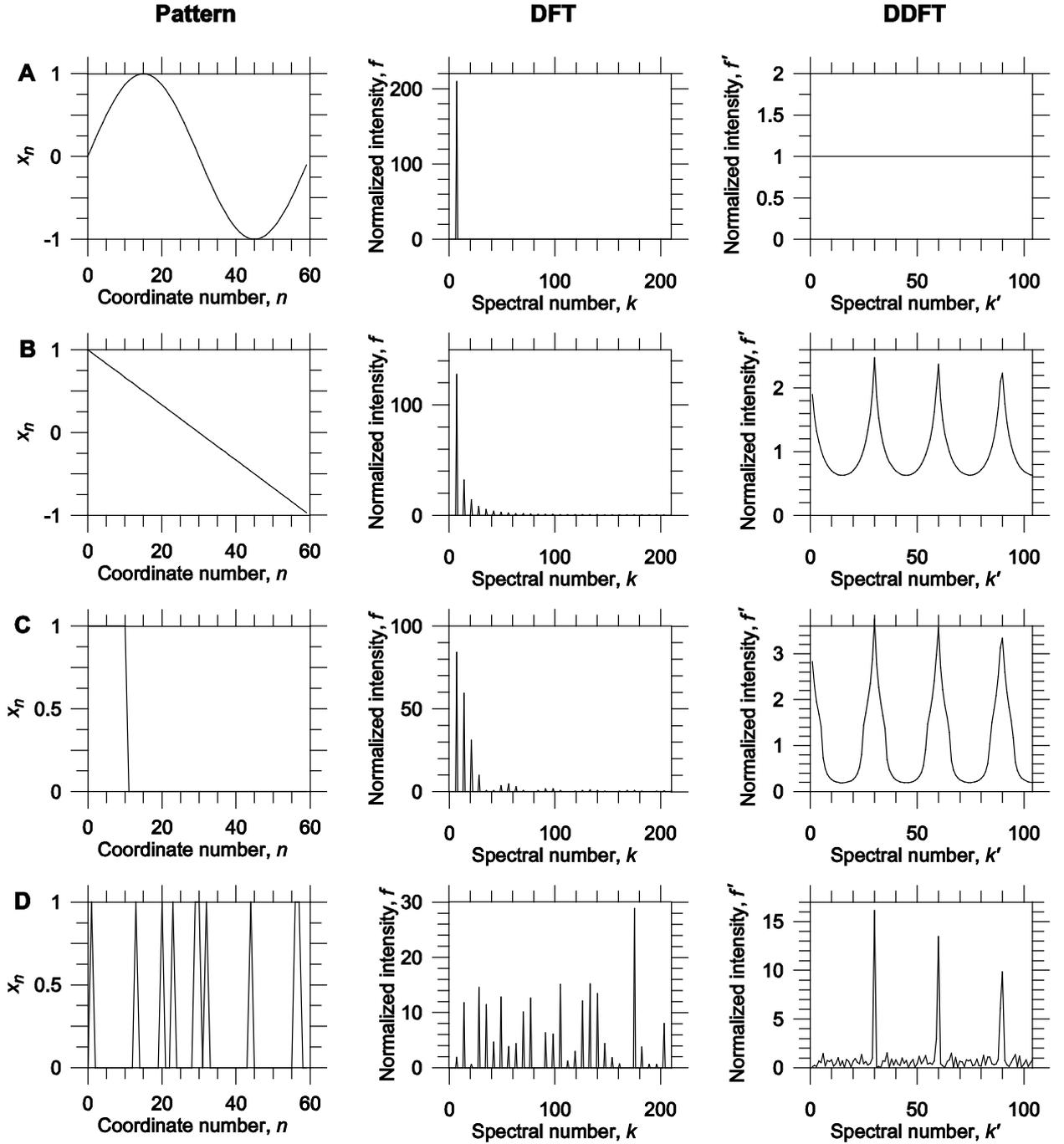

Fig. 1. DFT and DDFT spectra for various periodic patterns. A, $x_n = \sin(2\pi n/P)$, $P = 60$, $n = 0, 1, ..., 59$; B, $x_n = (30 - n)/30$; $n = 0, 1, ..., 59$; C, $x_n = 1$, if $n = 0, 1, ..., 10$ and $x_n = 0$, if $n = 11, 12, ..., 59$; and D, sequence of pulses modeled by Boolean variables taking the values 0 and 1, which are randomly distributed within the interval $n = 0, 1, ..., 59$. The each periodic pattern was repeated 7 times.

### 2.2.3. Generalizations of DDFT

Before calculating DDFT, the intensity harmonics (11) can be subjected to any other transform retaining ranking of their heights, $f_k \rightarrow \Phi(f_k)$, where the function $\Phi(f)$ should be monotonically increasing. In particular, such property is fulfilled for the pure power, Box-Cox [37, 38], Yeo–Johnson [39], and logarithmic transforms widely used in statistical data analysis. In the case of logarithmic transform, $f_k \rightarrow \log f_k$, the iterated Fourier transform is reduced to the well-known cepstrum transform (for a review see, e.g., [40, 41] and references therein). In particular, the cepstrum transform is suitable for processing of data with convolutions, superposition of direct and retarded signals, and signals with reverberations. Note that the cepstrum transform uses the inverse Fourier transform for $\log f_k$ (that determines the terminology in this technique), whereas in DDFT and its generalizations we use the forward Fourier transform for the spectral intensity harmonics or for the functions of them (that determines the term "double Fourier transform"). Such difference in definitions is not essential for Fourier transforms of real-valued data.

The preliminary mappings of intensity Fourier harmonics provide additional efficient tools for data preprocessing. The choice of optimal spectral mapping $f_k \to \Phi(f_k)$ before the second Fourier transform is highly problem and data dependent. It depends also on the desired balance between sensitivity and robustness of detection technique. Return again to the simple example with strict periodic patterns considered in Section 2.2.1. The intensity harmonics (15) in this case are combined by the product of the factor $\Phi_{k;P}$, which is strictly periodic on $k$ (Eq. (17)), and the non-periodic on $k$ factor $F_{k;P}$. The direct application of the cepstrum transform to $F_k = \Phi_{k;P} F_{k;P}$ is impossible because the factor $\Phi_{k;P}$ is equal to zero for the spectral numbers $k$ different from $k = N/P, 2N/P, ..., (P-1)N/P$. Either the additional filtering of the negative values of $\log f_k$ or the regularization $f_k \to f_k + 1 \to \log(f_k + 1)$ are needed before the second Fourier transform in this example. In Rayleigh spectra the characteristic maximum and minimum values of normalized harmonics are approximately $f_{\max} \approx \ln N$, $f_{\min} \approx 1/N$ [8]. Upon mapping $f_k \to \log f_k$, the negative outbursts in $\log f_k$ spectra would be about the positive outbursts in $f_k$ spectra.

### 2.2.4. Why DDFT?

The advantages in application of DDFT (or its generalizations described in the section above) to search for large-scale multi-periodic patterns are as follows. (i) The multiple equidistant series in DFT spectra are mapped onto the strongly reduced equidistant series in DDFT spectra. In this case DDFT compresses the information and facilitates the visualization and analysis of large-scale noisy periodicities. (ii) The high peaks in the range of low spectral numbers in DFT spectra which can be potentially related to the large-scale periodicities are often superimposed on peaks from the non-periodic sources. The strong trend in this range hampers the correct statistical assessment of observable effects. DDFT filters out the effects unrelated to the periodicities and maps large-scale periodicities onto the range of high spectral numbers without trend or with much weaker trend. (iii) Previously, it was suggested to identify noisy multi-periodic patterns via the sums of equidistant harmonics surrounded by variable windows [8, 35]. The number of equidistant harmonics and the width of windows vary in the different sums and affect the relevant statistical criteria for the assessment of significance of observed periodicity. DDFT provides a unified and more convenient approach to the study of equidistant series in DFT spectra.

Although the application of DDFT is formally not restricted, the technique is efficient if the value of period $p$ exceeds the total number of periods $N_p$ in the data set

$$p > N_p \text{ or } \sqrt{N} > k_g \tag{27}$$

Fig. 2 illustrates the situation when $p < N_p$. In this case the peaks in DDFT spectra turn out to be more frequent than those in the primary DFT spectra. The subsequent iteration of DDFT would reproduce all the difficulties related to the primary DFT. The range in DFT spectrum with the spectral numbers corresponding to the generating harmonics obeying inequality (27) $k_g \in (2, \sqrt{N})$ is mapped by DDFT onto $k'_{II} \in (\sqrt{N}/2, K'_{II})$. Primarily, DDFT technique is aimed at the analysis of hidden periodicities within these ranges.

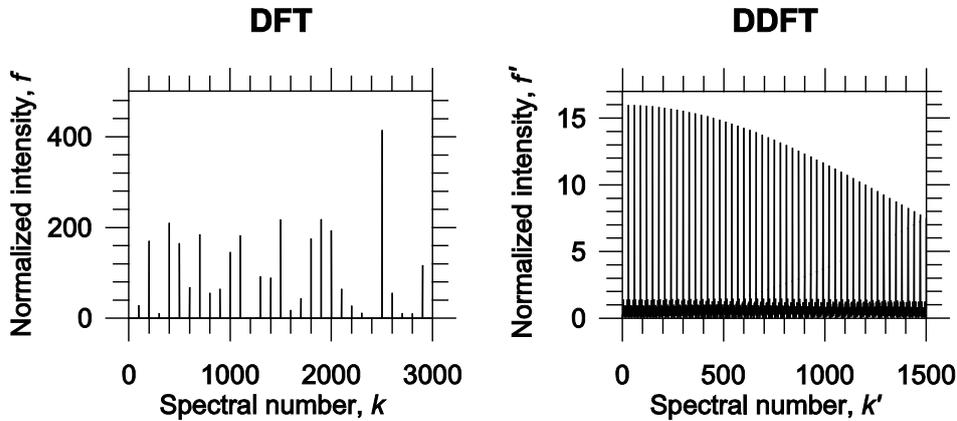

Fig. 2. DFT (left) and DDFT (right) spectra for the repeating sequence of pulses modeled by Boolean variables taking the values 0 and 1 randomly distributed within the interval $n = 0, 1, ..., 59$. The periodic pattern was the same as in Fig. 1D and was repeated 100 times.

## 2.3. Statistical criteria

### 2.3.1. Gaussian white noise and Rayleigh spectra

In the noisy spectra the heights of harmonics (11) or (22) should be statistically assessed. In this section we present some statistical criteria concerning this problem. It is known that Fourier transform of white noise provides the distribution of harmonics obeying Rayleigh statistics [42, 43]. We reproduce this proof for completeness because the similar technique is used below for the consideration of DDFT statistics. Let the variables $\{\Delta x_n\}$ be independent and obey the identical Gaussian distribution. Then, the multi-variable probability density for the complete set of variations $\{\Delta x_n\}$ will be given by the product

$$p(\Delta x_0; ...; \Delta x_{N-1}) = \prod_{n=0}^{N-1} p(\Delta x_n) \qquad (28)$$

where

$$p(\Delta x) = \frac{1}{(2\pi\sigma^2)^{1/2}} \exp(-\Delta x^2 / 2\sigma^2) \qquad (29)$$

The averaging of the spectral intensity (6) over distribution (28) provides the ergodic relationship valid for $N \gg 1$, because the estimated variance (10) converges to the fixed value $\sigma^2$ at this limit,

$$\overline{F} = <F> = \sigma^2 \qquad (30)$$

where the mean $\overline{F}$ is defined by Eq. (10), while the angular brackets correspond to the averaging over distribution (28) (or to the averaging over ensemble of random realizations). DFT of real independent Gaussian variables $\{\Delta x_n\}$ yields spectral intensity obeying Rayleigh distribution.

The proof is straightforward. For real random variables $\{\Delta x_n\}$ the generating function for Fourier harmonics can be conveniently defined as (see, e.g., [42–44])

$$<Z> = <\exp\left(i\sum_{k=1}^{K} u_k X_k + i\sum_{k=1}^{K} u_k^* X_k^*\right)> \qquad (31)$$

where the auxiliary variables $\{u_k\}$ and the conjugate variables $\{u_k^*\}$ are assumed to be independent and the angular brackets denote averaging over $\{\Delta x_n\}$. The various spectral moments can be obtained by differentiating the generating function (31) with respect to the auxiliary variables $\{u_k\}$ and $\{u_k^*\}$,

$$<X_{j_1}...X_{j_l} X_{k_1}^*...X_{k_m}^*> = \frac{\partial^{l+m}}{i^{l+m} \partial u_{j_1}...\partial u_{j_l} \partial u_{k_1}^*...\partial u_{k_m}^*} <Z> \Bigg|_{\{u_k\}, \{u_k^*\}=0} \qquad (32)$$

The direct calculations for the independent Gaussian variables $\{\Delta x_n\}$ provide the related generating function

$$<Z> = <\exp\left(i\sum_{k=1}^{K} u_k X_k + i\sum_{k=1}^{K} u_k^* X_k^*\right)> =$$
$$<\exp\left(iN^{-1/2}\sum_{k=1}^{K} u_k \sum_{n=0}^{N-1}\Delta x_n e^{-i2\pi nk/N} + iN^{-1/2}\sum_{k=1}^{K} u_k^* \sum_{n=0}^{N-1}\Delta x_n e^{i2\pi nk/N}\right)> =$$
$$<\exp\left(iN^{-1/2}\sum_{n=0}^{N-1}\Delta x_n \left(\sum_{k=1}^{K} u_k e^{-i2\pi nk/N} + \sum_{k=1}^{K} u_k^* e^{i2\pi nk/N}\right)\right)> =$$
$$\exp\left(-\frac{\sigma^2}{2N}\sum_{n=0}^{N-1}\left(\sum_{k=1}^{K} u_k e^{-i2\pi nk/N} + \sum_{k=1}^{K} u_k^* e^{i2\pi nk/N}\right)^2\right) = \exp\left(-\sigma^2 \sum_{k=1}^{K} u_k u_k^*\right) \quad (33)$$

The generating function (33) corresponds to the multi-variable Rayleigh probability for the intensity harmonics (6)

$$\Pr(F_1 > F_1'; \ldots; F_K > F_K') = \exp\left(-(F_1' + \ldots + F_K')/\sigma^2\right) \quad (34)$$

As for $N \gg 1$ the estimated variance (see Eq. (10)) converges to the exact $\sigma^2$ for the Gaussian probability (29), the distribution of normalized harmonics (11) can be described in terms of universal multi-variable probability independent of $\sigma^2$

$$\Pr(f_1 > f_1'; \ldots; f_K > f_K') = \exp(-f_1' - \ldots - f_K') \quad (35)$$

This proves that universality of the Rayleigh distribution in spectral problems is at least the same as universality of the Gaussian distribution for the random real variables. The relationship between the Gauss and Rayleigh distributions indicates that in the presence of noise the periodic features should be retained in more than $2\sqrt{N}$ positions coordinately distributed over the data set to be detected by the peaks in DFT spectrum.

In the problems with the strictly fixed variance (10) (or in the problems with the strict sum rule for the intensity harmonics), the Rayleigh distribution (35) should be replaced by De Finetti distribution [44]

$$\Pr(f_1 > f_1'; \ldots; f_K > f_K') = \left(1 - f_1'/K - \ldots - f_K'/K\right)_+^{K-1} \quad (36)$$

where $x_+ = x$, if $x > 0$ and $x_+ = 0$, if $x \leq 0$. De Finetti distribution tends asymptotically to the Rayleigh distribution for $K \gg 1$. Some calculations of different characteristics related to De Finetti distribution can be found in Ref. [45]. De Finetti distribution should also be used, when the spectrum under study is compared with the reference random distribution having the same observable variance (10).

If random noise is distributed by the Gaussian probability (29) and the underlying patterns are composed of the strict repeats of a period $P$, the generating function is equal to

$$<Z> = \exp\left( i\sum_{k=1}^{K} u_k \chi_{k;P} X_{k;P} + i\sum_{k=1}^{K} u_k^* \chi_{k;P}^* X_{k;P}^* - \sigma^2 \sum_{k=1}^{K} u_k u_k^* \right) \quad (37)$$

(see Eq. (14) for nomenclature). The differentiation of $<Z>$ with respect to the auxiliary variables $\{u_k\}$ and $\{u_k^*\}$ provides the mean spectral intensity of the form

$$<F_k> = \Phi_{k;P} F_{k;P} + \sigma^2 \quad (38)$$

where the first term in r.h.s. is defined by Eqs. (15) and (16). The corresponding variance for intensity harmonics is

$$\sigma^2(F_k) = <F_k^2> - <F_k>^2 = 2\Phi_{k;P} F_{k;P} \sigma^2 + \sigma^4 \quad (39)$$

*2.3.2. Statistics for DDFT spectra*

Consider now the distribution of harmonics (19) corresponding to DDFT of Fourier harmonics $\{f_k\}$ obeying the probability (35). The corresponding mean spectral intensity (23) can be assessed via the asymptotic ergodic relationship (30),

$$<F_{II}> = <(f - <f>)^2> = 1 \cong \overline{F}_{II} \quad (40)$$

Therefore, the distributions for harmonics (19) and (22) are asymptotically identical, if the normalized harmonics (11) are distributed by the Rayleigh probability (35). The calculations similar to those above provide the generating function for DDFT harmonics

$$<Z_{II}> = 1/\prod_{k=0}^{K-2}\left( 1 - i(K-1)^{-1/2}\left( \sum_{k'=1}^{K'_{II}} u_{k'} e^{-i2\pi k'k/(K-1)} + \sum_{k'=1}^{K'_{II}} u_{k'}^* e^{i2\pi k'k/(K-1)} \right) \right) \quad (41)$$

Using the cumulant function and expanding the logarithms into Taylor series yields at $K \gg 1$ the asymptotic convergence to the cumulant function corresponding to the universal multi-variable Rayleigh probability (35),

$$\ln <Z_{II}> = -\sum_{k=0}^{K-2} \ln\left(1 - i(K-1)^{-1/2}\left(\sum_{k'=1}^{K'_{II}} u_{k'} e^{-i2\pi k'k/(K-1)} + \sum_{k'=1}^{K'_{II}} u_{k'}^* e^{i2\pi k'k/(K-1)}\right)\right) \approx$$
$$-\sum_{k'=1}^{K'_{II}} u_{k'} u_{k'}^* + O(1/K^{1/2}) \qquad (42)$$

This conclusion is in agreement with the results of numerical simulations (see Fig. 3 below). The notation $O(1/K^{1/2})$ means that all higher cumulants are small on the powers of parameter $1/K^{1/2}$. The correlations between harmonics with $k \neq k'$ calculated with the cumulant function (42) are

$$<f_{k;II} f_{k';II}> - <f_{k;II}><f_{k';II}> = 6/(K-1) \cong O(1/K) \qquad (43)$$

i.e., the correlations are small if $K >> 1$. The same concerns the cumulant

$$<f_{k;II}^2> - 2<f_{k;II}>^2 = 6/(K-1) \cong O(1/K) \qquad (44)$$

The estimates also show that in the presence of noise the equidistant features should be retained for more than $2\sqrt{K}$ harmonics in DFT equidistant series to be detected by peaks in DDFT spectrum.

The asymptotic expression for the cumulant function (42) proves that the heights of normalized noisy DDFT intensity harmonics may also be assessed by universal Rayleigh distribution. Fig. 3 shows the empirical distributions of the intensity harmonics for DFT of Gaussian variables $x_n$ and for related DDFT. The statistical significance of the heights of harmonics in the whole spectrum or in a range of the spectrum should be assessed by the extreme value statistics [8, 34, 46].

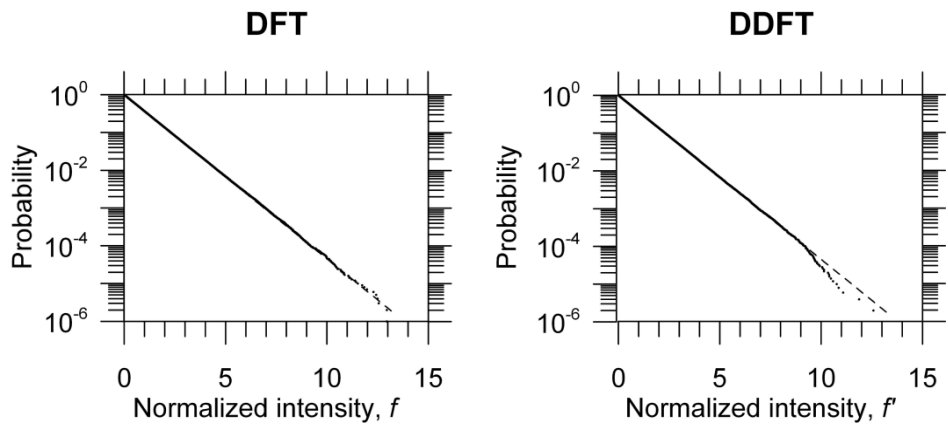

Fig. 3. The empirical probabilities for DFT spectrum (left) and for related DDFT spectrum (right). The empirical probability distributions are defined as $P(f) = K(f' > f)/K$, where $K(f' > f)$ is the number of normalized intensity harmonics in DFT or DDFT spectra exceeding the threshold $f$ and $K$ is the total number of harmonics in the half of symmetrical Fourier spectra. The initial data correspond to a set of $2\times10^6$ Gaussian random deviations $\{\Delta x_n\}$ with the zero mean and variance $\sigma^2 = 1$.

In the state-of-the-art studies, the data are often analyzed by parallel multi-channel processing of different characteristics. Below we present some results for this case as well. Consider the sum of $r$ independent identically distributed harmonics obeying the universal Rayleigh statistics (35),

$$S_{k,r} = f_k^{(1)} + f_k^{(2)} + ... + f_k^{(r)} \tag{45}$$

Upon DDFT of this sums, the mean intensity in the resulting DDFT spectrum can be assessed as

$$<F_{II,S_r}> = <(S_r - <S_r>)^2> = r \cong \overline{F}_{II,S_r} \tag{46}$$

The calculations similar to those presented above provide the cumulant function for the normalized DDFT harmonics related to the sum (45)

$$\ln <Z_{II,S_r}> =$$
$$-\sum_{k=0}^{K-2} r \ln\left(1 - i(K-1)^{-1/2}\left(\sum_{k'=1}^{K'_{II}} (u_{k'}/r^{1/2}) e^{-i2\pi k'k/(K-1)} + \sum_{k'=1}^{K'_{II}} (u_{k'}^*/r^{1/2}) e^{i2\pi k'k/(K-1)}\right)\right) \approx$$
$$-\sum_{k'=1}^{K'_{II}} u_{k'} u_{k'}^* + O\left(1/(rK)^{1/2}\right) \tag{47}$$

i.e. the statistics for resulting normalized DDFT harmonics related to the sums (45) again obeys asymptotically Rayleigh distribution. The convergence to Rayleigh statistics becomes faster with the increase of $r$.

### 2.3.3. Method of periodograms

Both DFT and DDFT spectra (or counterpart spectral ranges) can be averaged over multiple noisy realizations (method of periodograms [12]). If the number of averaged spectra is large, $N_s \gg 1$, the random counterparts of intensity harmonics (11) or (22) in averaged spectral ranges can approximately be described by Gaussian statistics with the mean 1 and the variance $1/N_s$. The resulting averaged spectral ranges can be analyzed in terms of normalized deviations

$$\varsigma_k = \left(<f_k>_{spectra} - 1\right)/(1/N_s)^{1/2} \tag{48}$$

where $<f_k>_{spectra}$ is the $k$-th harmonic (11) or (22) averaged over spectra (or over counterpart spectral ranges). If the averaged random spectra are used as a reference for the periodogram under study, the normalized deviations (48) provide convenient variables for the assessment of their statistical significance. The significance of deviations (48) in a spectral range should be assessed by the extreme value statistics for Gaussian variables.

## *2.4. Large-scale periodicities under the action of noise with long-range correlations*

To illustrate why the noise with long-range variations is especially detrimental for the detection of large-scale multi-periodic patterns, we consider the white noise smoothed over sliding window of width $w$, $\tilde{x}_n^{(w)} = (x_n + ... + x_{n+w-1})/w$, where $x_n$ correspond to Gaussian variables obeying distribution (29). To avoid the end effects, the variables are circularly continued

$$x_n^{(c)} = \begin{cases} x_n, \text{ if } n < N \\ x_{n-N}, \text{ if } n \geq N \end{cases} \qquad (49)$$

The DFT generating function for the Gaussian smoothed and circularly continued variables is given by

$$<Z_w> = \exp\left(-(\sigma^2/w^2)\sum_{k=1}^{K}\omega_k\omega_k^* u_k u_k^*\right) \qquad (50)$$

where

$$\omega_k = \sum_{m=0}^{w-1} e^{i2\pi mk/N} = (1-e^{i2\pi kw/N})/(1-e^{i2\pi k/N}) \qquad (51)$$

The corresponding mean normalized intensity harmonics are determined as

$$<f_k^{(w)}> = (N-1)\sin^2(\pi kw/N)/w(N-w)\sin^2(\pi k/N) \qquad (52)$$

while the resulting probability distribution for the normalized harmonics is

$$\Pr(f_1 > f_1'; ...; f_K > f_K') = \exp(-f_1'/<f_1^{(w)}> - ... - f_K'/<f_K^{(w)}>) \qquad (53)$$

If the window width $w$ is commensurate with the total number of data points $N$, the normalized intensity harmonics (52) may be equal to zero at some spectral numbers $k$. In this case the harmonics (52) should be replaced by small finite values for the regularization of the probability (53). As is seen from Eq. (53), all moments of a chosen harmonic $f_k$ are proportional to the powers of $<f_k^{(w)}>$ ($<f_k^m> = m!<f_k^{(w)}>^m$). This means that both mean value and the fluctuations around it grow with the decrease of $k$ for small spectral numbers $k$ due to the growth of the normalized intensity harmonics $<f_k^{(w)}>$. As the peaks for large-scale multi-periodic patterns may be superimposed on that from the large-scale noisy variations in this range, such situation is especially detrimental for their detection.

The particular example of spectra for periodic patterns in the presence of noise with long-range correlations is shown in Fig. 4. As is seen from this figure, the equidistant peaks related to underlying periodicity are not discernible in DFT spectra in the vicinity of generating (or fundamental) harmonic (Fig. 4B, left), whereas they are clearly detectable in DDFT spectrum (Fig. 4B, right). The further details and results for this example can be found in Supplement.

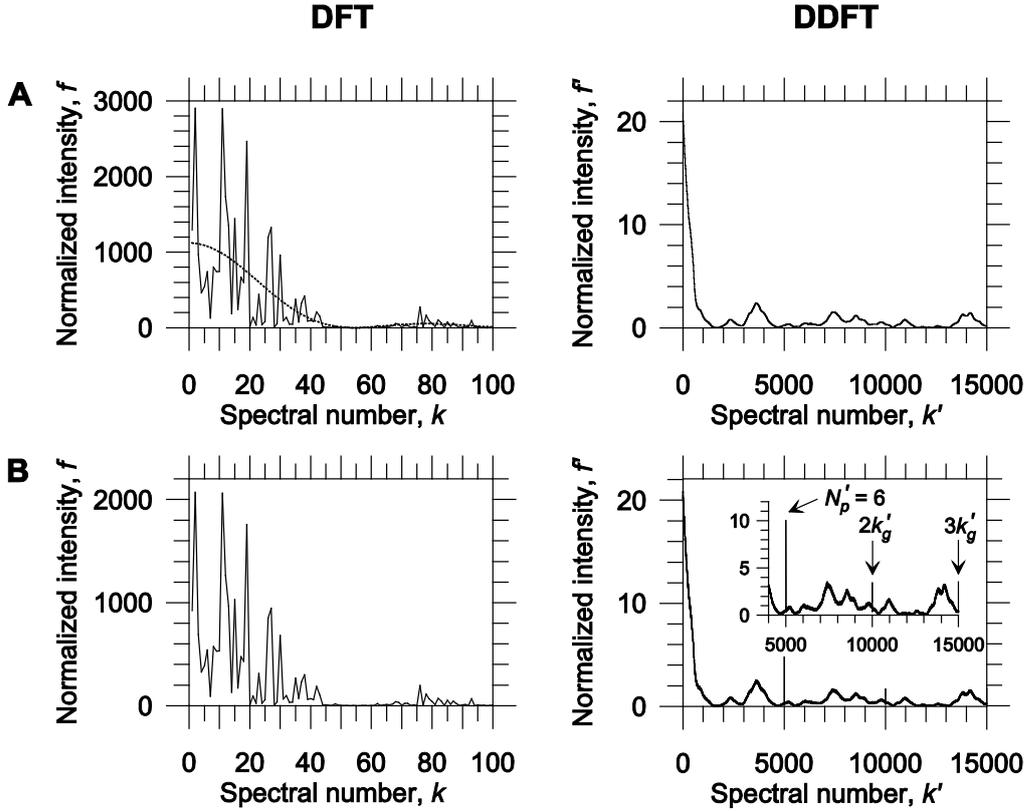

Fig. 4. The effect of noise with long-range correlations on the detection of large-scale periodic patterns. The repeating sequence of pulses was modeled by Boolean variables taking the values 0 and 1 randomly distributed within the interval of length $10^4$. The fraction of units was 0.1. The each periodic pattern was repeated 6 times. The correlated noise was generated by the smoothing of white noise within the sliding window of width $w = 1.1 \times 10^3$. A, The initial range of DFT spectrum (left) and DDFT spectrum (right) for a particular realization of correlated noise. The dotted line in DFT range shows the mean intensities over many realizations (Eq. (52)). B, The initial range of DFT spectrum (left) and DDFT spectrum (right) for the periodic patterns with a particular realization of correlated noise. The signal-to-noise ratio is slightly above the detection threshold, $r_{S/N} = 0.4$. The insert in the DDFT spectrum corresponds to the renormalized range (see Supplement for further details).

## 2.5. Concluding comments on DDFT

DDFT provides the efficient tool for filtering false positives and false negatives in the primary DFT spectra in comparison with commonly applied practice when the periodicities would be detected by the singular harmonics. The false positives in DFT spectra can be produced by the random outbursts or by the effects unrelated to periodic patterns (see the highest harmonics in DFT spectra shown in Figs. 1D and 4B), whereas the false negatives correspond to the missed patterns due to the seemingly low heights of generating harmonics associated with underlying periodicity (the harmonics with $k = 7$ and 6 in the same spectra). Even approximate correspondence between peaks in DFT and DDFT spectra

$$k \approx N_p \approx N'_{p;II} \tag{54}$$

strongly enhances their statistical significance. Therefore, the combined application of DFT and DDFT ensures the additional cross-check in search for noisy multi-periodic patterns.

## 2.6. Addendum: DDFT and correlation functions

The spectral intensity harmonics can be expressed by the circular correlation functions through the Wiener-Khinchin relationship [8, 36]. The circular correlation functions are defined as

$$K^c(m_0) = N^{-1} \sum_{n=0}^{N-1} x_n^c x_{n+m_0}^c, \quad m_0 = 0, 1, ..., N-1 \tag{A.1}$$

where

$$x_n^c = \begin{cases} x_n, & \text{if } 0 \leq n \leq N-1 \\ x_{n-N}, & \text{if } N \leq n \leq 2N-1 \end{cases} \tag{A.2}$$

The correlation functions are symmetrical relative to the middle,

$$K^c(m_0) = K^c(N - m_0) \tag{A.3}$$

The circular correlation functions (A.1) and the intensity harmonics (6) are related by the Wiener-Khinchin relationship,

$$F_k = \sum_{m_0=0}^{N-1} K^c(m_0) e^{i2\pi k m_0 / N} \tag{A.4}$$

as

$$N^{-1} \sum_{m_0=0}^{N-1} \sum_{n=0}^{N-1} x_n^c x_{n+m_0}^c e^{i2\pi k m_0/N} = N^{-1} \sum_{n=0}^{N-1} \sum_{m_0=0}^{N-1} x_n^c e^{-i2\pi k n/N} x_{n+m_0}^c e^{i2\pi k(n+m_0)/N} = X_k X_k^* = F_k$$

The inverse Fourier transform of (A.4) yields

$$K^c(m_0) = N^{-1} \sum_{k=0}^{N-1} F_k e^{-i2\pi k m_0 / N} \tag{A.5}$$

Substitution of (A.4) into Eq. (19) provides

$$X_{k';II} = \overline{F}^{-1}(K-1)^{-1/2} \sum_{k=0}^{K-2} \sum_{m_0=0}^{N-1} K^c(m_0) e^{i2\pi(k+2)m_0/N} e^{-i2\pi kk'/(K-1)}, \quad k' = 0, 1, ..., K-2 \tag{A.6}$$

Taking into account the relationships (25) and (26), Eq. (A.6) can also be presented in the form

$$X_{k';II} = \overline{F}^{-1}(K-1)^{-1/2} \sum_{m_0=0}^{N-1} K^c(m_0) e^{i4\pi m_0/N} \chi_{k';m_0} \tag{A.7}$$

where

$$\chi_{k';m_0} = \sum_{k=0}^{K-2} e^{i2\pi k(m_0/p'_{II} - k'/k'_{II,g})/N'_{p;II}} = \frac{1 - e^{i2\pi(K-1)(m_0/p'_{II} - k'/k'_{II,g})/N'_{p;II}}}{1 - e^{i2\pi(m_0/p'_{II} - k'/k'_{II,g})/N'_{p;II}}} \tag{A.8}$$

The factor $\chi_{k';m_0}$ tends to $K-1$ when the expression in the exponents tends to zero. This occurs at $k'/k'_{II,g} \approx 1$ and $m_0/p'_{II} \approx 1$; $k'/k'_{II,g} \approx 2$ and $m_0/p'_{II} \approx 2$ etc. Though the noisy periodicities may be not seen in the direct spectra for correlation functions, they can be displayed by transform (A.7).

## 3. Simulations

The details of simulations and relevant processing schemes are presented in Supplement. Here, we restrict ourselves to the main conclusions on the simulation results. They can be summarized as follows.

● Peaks in DDFT spectra are robust with respect to noise.

● DDFT is able to detect large-scale multi-periodic patterns with both integer and fractional number of patterns within interval under analysis. The superposition of both integer and fractional patterns can also be resolved by DDFT.

● DDFT is able to detect large-scale multi-periodic modulations of short-periodic signals. In the presence of short-periodic signals the efficient application of DDFT needs additional preprocessing of primary DFT spectra (cut-off of short-periodic peaks).

● At applying periodogram technique, the processing mode (i) DFT → averaging → DDFT is preferable over (ii) DFT → DDFT → averaging, i.e. in (i), first, DFT is calculated, then, DFT spectra are averaged over random realizations and, finally, DDFT spectrum is calculated for the resulting averaged DFT spectrum, whereas in (ii), first, DFT is calculated, then, DDFT and, finally, DDFT spectra are averaged over random realizations.

● DDFT is able to resolve between large-scale multi-periodic patterns and large-scale noisy variations. The de-trending of DFT spectrum in the range of low spectral numbers improves the detection of large-scale multi-periodic patterns via DDFT.

## 4. Examples

### 4.1. Solar rotation and variations in solar wind speed

The practical applications of DDFT technique can be illustrated with two following examples. The first one was taken from the solar-terrestrial physics. The experimental data were obtained by the Deep Space Climate Observatory (DSCOVR), which was launched on 11 February 2015. This spacecraft is orbiting the Sun-Earth Lagrange point 1 (the neutral gravity point L1) and monitoring the solar wind. The Faraday cup device on board DSCOVR provides measurements of solar wind parameters including the solar wind speed. The data are available for downloading at DSCOVR Space Weather Data Portal (URL https://www.ngdc.noaa.gov/dscovr/portal/index.html#/). The time interval analyzed in the present paper ranges from 26 July 2016 to 20 October 2018. One-minute averages were used in this study.

The data were preliminarily preprocessed by filtering out the outliers and filling the gaps by the box-car median averaging for 6-hour intervals. The gaps were filled by linear interpolation of the averaged data. The sampling time between consecutive points in the preprocessed data was 6 h. A plot for the time series obtained is shown in Fig. 5A. The equidistant series is clearly visible in related DFT spectrum shown in Fig. 5B. The distance between the equidistant peaks is slightly variable. The corresponding period estimated from the position of the first peak was 25.53 days. The DDFT spectrum (Fig. 5C) reveals a significant peak corresponding to the period of 27.02 days. The partial de-trending of the DFT spectrum in the range of low spectral numbers enhances the peak in DDFT spectrum (Fig. 5D) but retains the period.

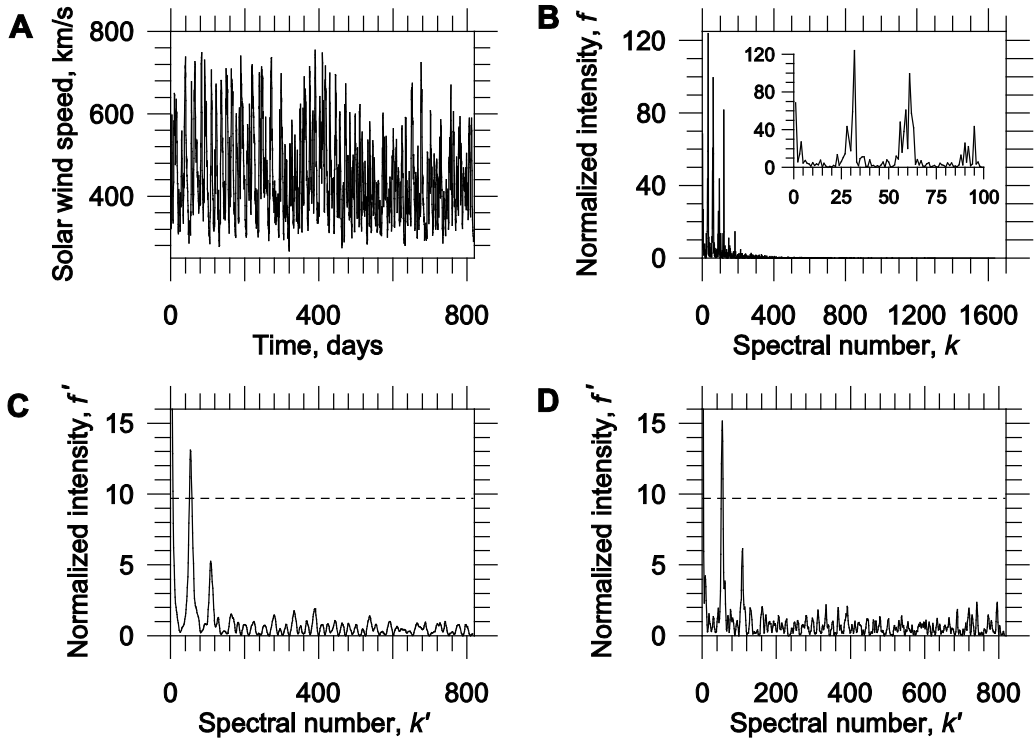

Fig. 5. The variations in solar wind speed. A, The data obtained by the Deep Space Climate Observatory from 26 July 2016 to 20 October 2018. B, The normalized DFT spectrum corresponding to data in the panel A. C, The related DDFT spectrum. D, DDFT spectrum after partial de-trending of DFT spectrum in the range of low spectral numbers. The dashed horizontal lines in the panels C and D mark the extreme value thresholds corresponding to the probability Pr = 0.05 for the Rayleigh spectra.

The period detected by DDFT is in good agreement with previous observations for rotation periods for different features in the solar atmosphere as observed from the Earth, i.e. synodic rotation periods (see for a review [47, 48] and references therein). In particular, the synodic rotation period obtained by Carrington from sunspot observations was 27.275 days, whereas the synodic rotation period obtained by Bartels [49] from tracking certain recurring or shifting geomagnetic features related to solar activity was 27.0 days. As is seen, the period obtained by DDFT is closer to the previous observations of the synodic rotation period in comparison with that obtained by DFT.

*4.2. Symbolic sequences and bioinformatics*

The second example refers to symbolic sequences. A wide field of symbolic sequences comprises a variety of letter texts, communication via dot-and-dash signals, symbolic dynamics [50, 51], substitution sequences [52], etc. However, presumably, the most important field of their application is related to the bioinformatic analysis of DNA/RNA sequences. In this section we briefly review the results obtained previously by the present authors [34, 53].

Let a symbolic sequence of length *M* be composed of letters (A, B, C, ...) positioned in the sites $m = 1, 2, ..., M$. Then, DFT of symbolic sequence is defined as

$$\rho_\alpha(q_k) = M^{-1/2} \sum_{m=1}^{M} \rho_{m,\alpha} e^{-iq_k m}, \quad q_k = 2\pi k / M, \quad k = 0, 1, ..., M-1 \tag{55}$$

where $\rho_{m,\alpha}$ indicates the position occupied by the symbol of type $\alpha$; $\rho_{m,\alpha} = 1$ if the symbol of type $\alpha$ occupies the $m$-th site and 0 otherwise. Fourier harmonics (55) are calculated for the symbol of each type separately. The other definitions and properties follow those presented in Section 2.1,

$$F_{\alpha\alpha}(q_k) = \rho_\alpha(q_k) \rho_\alpha^*(q_k) \tag{56}$$

The intensity harmonics (56) are often called the structure factors in the analysis of symbolic sequences. Their normalization is determined by the exact sum rule (see, e.g., [8, 36])

$$f_{\alpha\alpha}(q_k) = F_{\alpha\alpha}(q_k) / \overline{F}_{\alpha\alpha}; \quad \overline{F}_{\alpha\alpha} = N_\alpha (M - N_\alpha) / M(M-1) \tag{57}$$

where $N_\alpha$ is the total number of symbols of type $\alpha$ in a sequence of length $M$. The normalization makes the harmonics (57) to be independent of the relative occurrences of different symbols and unifies their statistical properties for the reference random sequences with the same composition of symbols [8, 36]. The combinations of different symbols can be conveniently studied with the sums over a part of chosen symbols

$$S(q_k) = \sum_\alpha f_{\alpha\alpha}(q_k) \tag{58}$$

The statistics of DDFT spectra related to the sums (58) obeys approximately Rayleigh distribution for the random sequences (see [34] and Eq. (47)). In DNA/RNA sequences the symbols correspond to the nucleotides of four types, $\alpha \in (A, C, G, T/U)$. As the coding in symbolic sequences differs strongly from sin/cos, the application of DDFT is especially efficient in this case.

The genomic DNA of a living organism is compacted into a chromosome, an organized multilevel hierarchical structure. Combining imaging techniques and high-throughput genomic mapping tools yield the information on chromatin interactions and on the chromosome architecture [54–56]. The chromosome folding consists of approximately repeating structural units at each structural level. As DNA compaction into chromosome is determined by its specific interactions with structural proteins, the related motifs should also be distributed multi-periodically over chromosome and can be detected by DFT and DDFT. These multi-periodic features ought to be detected on the strong noisy background induced by random point mutations and insertions/deletions during molecular evolution.

The combined DFT and DDFT technique was applied to search for large-scale multi-periodic patterns in the genome of bacterium *Escherichia coli* [34]. The genome of *E. coli* strain K-12 substrain MG1655 (GenBank assembly accession GCA_000005845.2) contains 4,641,652 base pairs and codes for more than 4,200 genes. The genetic and structural studies as well as the gene expression profiles revealed that the longest patterns can be associated with six domains, four macrodomains, Ori, Ter, Left, and Right, and two so-called non-structured regions, NS-left and NS-right [57, 58]. In a writhed chromosome structure resembling a double-twisted thick ring (shown in Fig. 6A), the six domains are positioned approximately on semi-ellipses, which may also be considered as repeating structural units of the *E. coli* chromosome.

The resulting DFT spectra for the *E. coli* genome in the range of low spectral numbers is enriched by the high peaks due to the mosaic patchiness of the genome. The effects unrelated to the periodicity appear to be much stronger than the contributions from multi-periodic patterns. Therefore, the harmonic with the spectral number $k = 6$ is unremarkable in comparison with the other peaks in this range. On the contrary, DDFT spectrum in the range corresponding to the large-scale patterns reveals clearly the related underlying periodicity (Fig. 6B). The application of Morlet wavelet transform to search for large-scale genome regularities in the *E. coli* genome detected patterns corresponding to 7.1–7.7 domains in a scalogram [59], whereas DDFT detects 6.3 long periods (Fig. 6B) that is in the better agreement with the genetic and structural data. The resolution of periodic patterns by scalogram is also much broader in comparison with DDFT. Unlike DDFT spectrum, the application of autocorrelation functions cannot visually display the underlying large-scale multi-periodic variations (see the relevant spectra in [34]).

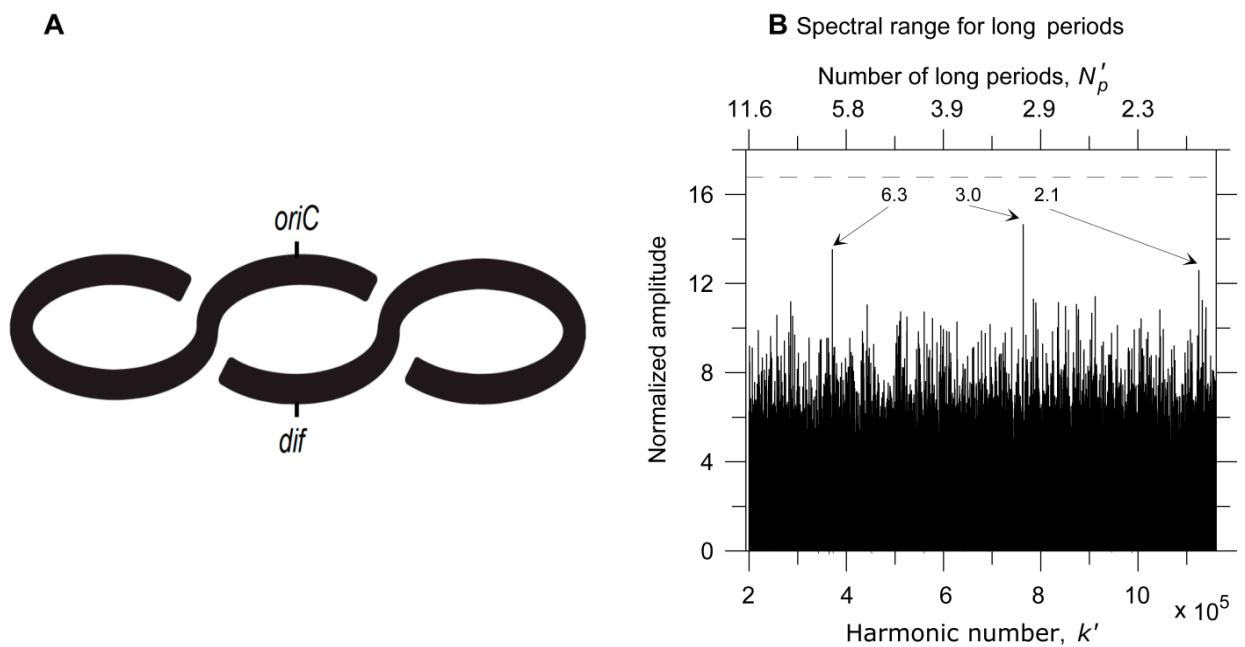

Fig. 6. The large-scale multi-periodic patterns in the *E. coli* genome and DDFT spectrum. DDFT was performed for the sum (58) over all four nucleotides. A, The schematic representation of the *E. coli* chromosome in rapidly growing bacteria. The locus *oriC* refers to the origin of chromosome replication, while *dif* refers to the site, where replicated chromosome dimers are resolved. B, The three highest harmonics in DDFT spectral range corresponding to the large-scale patterns are indicated by arrows and the numbers of corresponding long periods are shown explicitly. The horizontal dashed line refers to Pr = 0.05 probability threshold for the Rayleigh extreme value statistics in the chosen spectral range.

The similar ideas based on the multi-periodic distribution of motifs corresponding to specific DNA-protein interactions over the genome can be applied also to the analysis of genome packaging within capsids of viruses (protein envelopes of viruses) with icosahedral symmetry [53] (for a review on icosahedral viruses see, e.g., [60–62]). The corresponding multi-periodic segmentation in viral genomes should be coordinated with the (multiple) elements of icosahedral symmetry. Combining DFT and DDFT, we studied the quasi-regular large-scale segmentation in genomic sequences of different ssRNA, ssDNA, and dsDNA viruses and found the significant quasi-regular segmentation of genomic sequences related to the virion assembly and the genome packaging within icosahedral capsid. We also found good correspondence between our results and available cryo-electron microscopy data on capsid structures and genome packaging in these viruses. These results are of basic interest and may be helpful for the development of anti-viral drugs and targeted therapy. We envisage the enormous potential of DDFT applications to the analysis of DNA/RNA sequences. Though we picked up the bioinformatic examples for the spectral analysis of symbolic sequences, the approach remains universal for the general symbolic sequences.

### *4.3. Comparison: DDFT versus cepstrum*

We compared the results obtained by DDFT and by the original and modified cepstrum transform for the both problems considered above. As an illustration, we present, first, the variants of spectra obtained by the cepstrum technique for variations in solar wind speed (shown in Fig. 7; see also Figs. 5A and 5B for the input data). As is seen from the spectra shown in Fig. 7, the cepstrum transform "as it is" (first, the normalized intensity harmonics in DFT spectrum are mapped to $f_k \to \ln f_k$ and, then, the second Fourier transform is applied to $\ln f_k$) appears to be very sensitive to the harmonics with small heights and is not able to resolve the underlying periodicity (Fig. 7A). After filtering out the harmonics with small heights or after the additional regularization $f_k \to f_k + 1 \to \ln(f_k + 1)$, the corresponding peaks become clearly detectable in the resulting spectra (Figs. 7B and 7C, respectively). The spectral numbers for the peaks in the second Fourier spectra for both DDFT and the variants of the cepstrum transform are the same for all cases. Thus, the detected periodicity in variations of solar wind speed remains the same for all these techniques with the second Fourier transform.

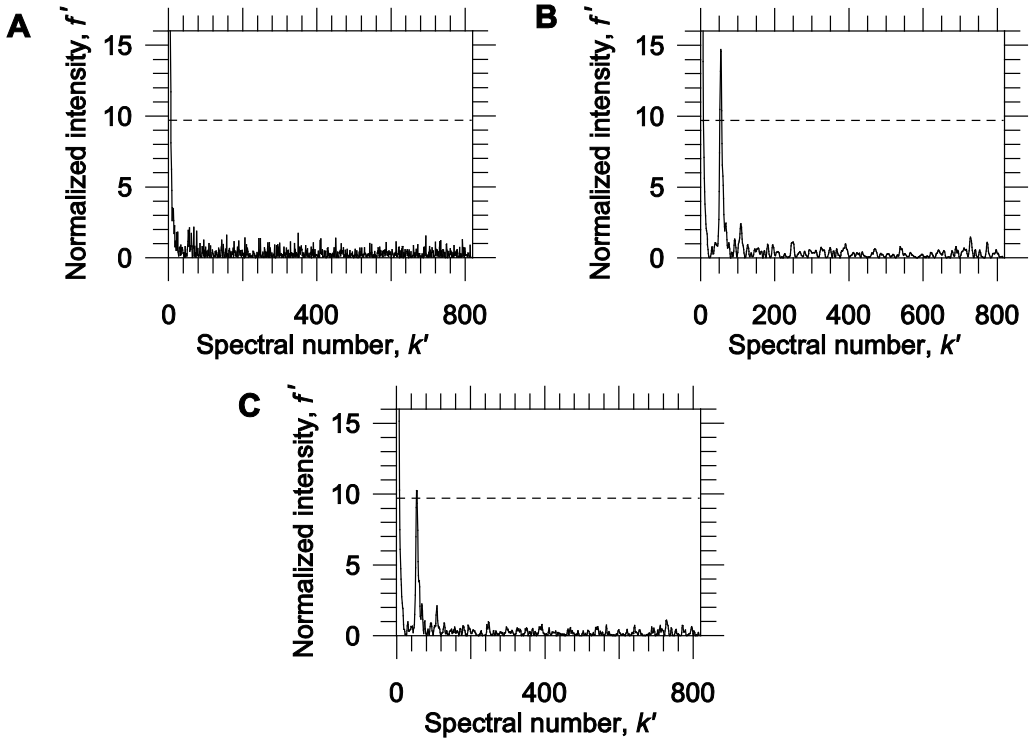

Fig. 7. The normalized spectra obtained by the cepstrum technique and its modifications for variations in solar wind speed. A, Cepstrum was used "as it is". First, the normalized intensity harmonics in DFT spectrum were mapped to $f_k \to \ln f_k$ and, then, the second Fourier transform was applied to $\ln f_k$. B, First, the normalized intensity harmonics in DFT spectrum were mapped to $f_k \to \ln f_k$, then, the harmonics with negative values of $\ln f_k$ were filtered out, and, finally, the second Fourier transform was applied to such filtered spectrum. C, The normalized intensity harmonics in DFT spectrum were mapped to $f_k \to f_k + 1 \to \ln(f_k + 1)$ and, then, the second Fourier transform was applied to $\ln(f_k + 1)$. The dashed horizontal lines in the panels A–C mark the extreme value thresholds corresponding to the probability Pr = 0.05 for the Rayleigh spectra.

We tried also various preprocessing schemes for the DDFT and cepstrum technique in application to the analysis of large-scale quasi-periodic segmentation in the *E. coli* genome. The best variants are presented in Fig. 8. Again, we found that DDFT provides better resolution of underlying large-scale periodicities in comparison with the cepstrum technique.

Primarily, the application of the cepstrum transform was motivated by the factorized form of Fourier spectral harmonics for a class of problems and the general theory of homomorphic (i.e., linear in a generalized sense) mappings [40, 41]. In the problem related to the search for large-scale periodic patterns, a factorized form of Fourier spectral harmonics is retained only for one-periodic patterns (cf. Eqs. (15)–(17)) and fails for a superposition of multi-periodic patterns. Our results prove that both arguments (factorization and homomorphic mappings) are not crucial for the exclusive application of the cepstrum transform. The similar results can be obtained by DDFT, simpler, with less number of operations, and often with better resolution of underlying large-scale multi-periodic patterns.

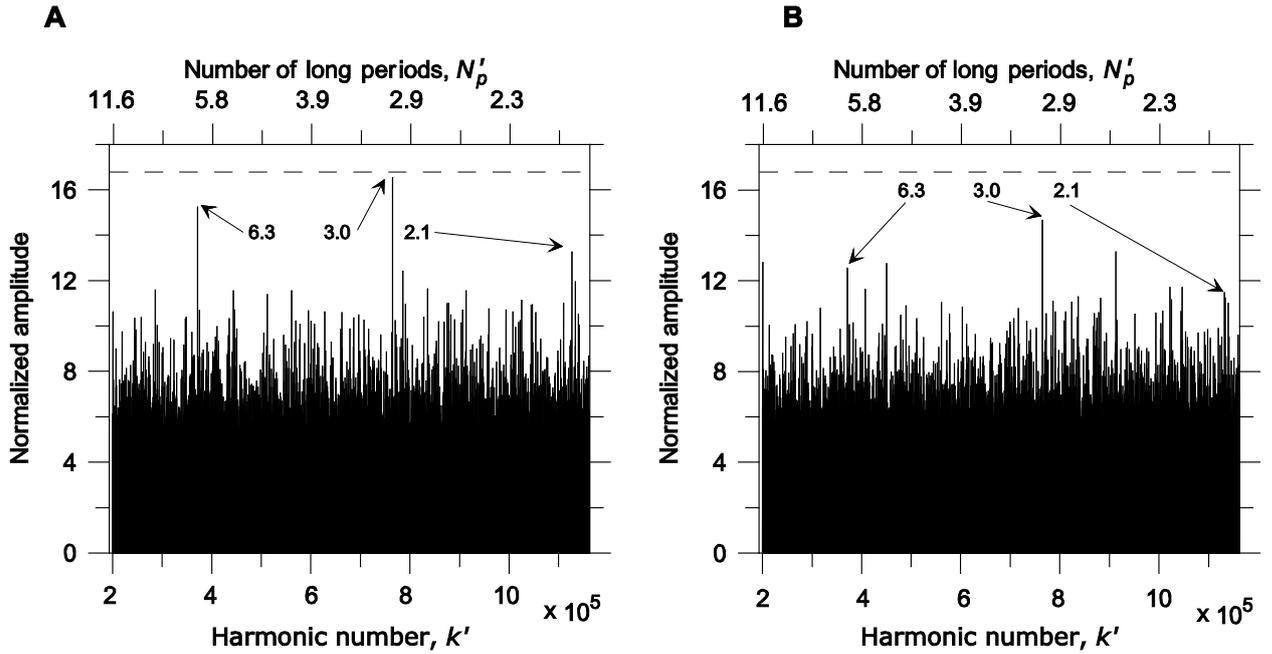

Fig. 8. The underlying large-scale periodic patterns resolved in the *E. coli* genome by the DDFT and cepstrum technique. A, Normalized DDFT spectrum obtained for the sum (58) over all four nucleotides. The harmonics with the heights less than unity were preliminarily filtered out before summation and implementation of DDFT. B, Normalized spectrum in the cepstrum technique for the sum over logarithms of the normalized intensity harmonics corresponding to the nucleotides of different types. All negative logarithms were preliminarily filtered out before the second Fourier transform. The horizontal dashed lines refer to Pr = 0.05 probability threshold for the Rayleigh extreme value statistics in the chosen spectral range.

## 5. Discussion

Commonly, the researchers restrict themselves only to the detection of fundamental frequency corresponding to the extreme left harmonic in the equidistant series or even consider the equidistance as undesirable effect needed to be specially suppressed. We tried to put this property on the regular and systematic ground especially useful for the detection of large-scale noisy multi-periodic patterns. DFT is reversible; there is one-to-one correspondence between data under analysis and DFT harmonics. Hence, the information is neither acquired nor lost during DFT. It merely is transformed. The information about multi-periodic patterns in underlying data is distributed over DFT spectrum strongly inhomogeneously. The harmonics for short-periodic patterns occupy a relatively broad range in DFT spectrum, whereas the harmonics for the fundamental periods of large-scale periodic patterns are squeezed in a relatively narrow range of low spectral numbers. DDFT inverts such relationship and in this sense appears to be dual to DFT (this is also the reason why the subsequent iterations of Fourier transform are inefficient). In the problems related to the search for large-scale noisy multi-periodic patterns, the characteristic corresponding to the equidistant series of harmonics dually displayed by DFT and DDFT spectra is the number of periodic patterns in underlying data. As in DDFT spectrum the fundamental harmonics for large-scale periodic patterns occupy now much broader range, the numbers of underlying large-scale periodic patterns can be resolved by DDFT with high accuracy even in the case of fractional periodic patterns (see Section 4 and Supplement). All operations in DDFT can be performed by standard packages for Fourier transform and statistics. This makes the DDFT technique be practical for various applications.

The search for periodic patterns implies their sequential ordering with time or length. In the problems with simultaneous processing of different signals (e.g., the composition of direct and retarded signals) the quasi-periodic variations in DFT intensity spectra may be of different nature (related to the retardation time in the mentioned example [40, 41]). Though the application of cepstrum transform to the latter problem looks natural, the trial of alternative techniques would be interesting even in this case (cf. also the results of our simulations on the co-existence of short and long periodic patterns in Supplement).

Our results prove that DDFT provides a powerful tool to search for large-scale noisy multi-periodic patterns. DDFT is able to distinguish multi-periodicities from large-scale non-periodic variations. The DDFT technique is efficient under two restrictions: (i) the shape of repeating patterns should be different from sin/cos components of DFT basis (this restriction is fulfilled for majority of patterns); and (ii) the number of sampling points within a repeat should exceed the total number of periods. To ensure the good statistics and the property of self-averaging, the total number of sampling points in the data set should be large enough. To unify the analysis and to include also the superposed sin/cos-like patterns into the whole scheme, we recommend to combine DFT and DDFT. Such a combination provides an additional cross-check for the detection of hidden multi-periodic patterns (see Eq. (54)). DDFT is able to resolve a superposition of patterns with incommensurate periods and also resolves the fractional patterns. Both these problems cannot be solved using autocorrelation functions without additional sophisticated preprocessing (cf. [34]). Unlike transform using Ramanujan sums, DDFT does not need the matrix inversion of large rank. In particular, the analysis of the *E. coli* genome with RPT would need the inversion of matrix $4{,}641{,}652 \times 4{,}641{,}652$, which is computationally an enormous task. The analytical criteria for the assessment of the harmonic heights in RPT spectra are also absent.

In many engineering, physical, astrophysical, medical, and economic problems monitoring of data is limited by the time or length. In such cases by necessity the measurements are restricted by several periods. However, in the majority of problems large-scale multi-periodic patterns are primarily interesting as reflecting the hierarchical organization (temporal and/or spatial) of a phenomenon or an object under study. We hope that DDFT can facilitate the data processing and provide the insight into these problems as well.

[57] C.J. Dorman, Genome architecture and global gene regulation in bacteria: making progress towards a unified model? *Nature Rev. Microbiol.* **11** (2013) 349–355.

[58] S.J. Messerschmidt and T. Waldminghaus, Dynamic organization: chromosome domains in *Escherichia coli*. *J. Mol. Microbiol. Biotechnol.* **24** (2014) 301–315.

[59] T.E. Allen, N.D. Price, A.R. Joyce and B.Ø. Palsson, Long-range periodic patterns in microbial genomes indicate significant multi-scale chromosomal organization, *PLoS Comput. Biol.* **2** (2006) e2.

[60]  R. Calendar and S.T. Abedon, (eds.), The Bacteriophages, 2nd Edition, Oxford Press, 2006.

[61] M.G. Rossmann and V.B. Rao, (eds.), Viral Molecular Machines, Springer, 2012.

[62] M.G. Mateu, (ed.), Structure and Physics of Viruses, Springer, 2013.
33

# Supplement

## Results of simulations

In this supplement we present the results of simulations to demonstrate the abilities of DDFT and its robustness to the action of noise. Throughout the simulations we will use linear superposition of patterns with different periods $p_1$ and $p_2$, $x'_n = k_1 x_n^{(p_1)} + k_2 x_n^{(p_2)}$. The noise was added to the signal as $x'_n = x_n^{(p_1)} + \xi_n$. The noise intensity varied to cover a range of signal-to-noise ratios and to determine the signal detection threshold. The statistical criteria for the threshold were described in Section 2.3. For the reasons of visualization we chose the spectra corresponding to $r_{S/N}$ slightly above the threshold. In Sections S.1–S.4 the noise is white, whereas in Section S.5 we consider the noise with the long-range correlations. The main aim of the consideration below consists in illustrating typical situations and know-how processing in search for large-scale quasi-periodic patterns in noisy data series with DDFT.

The standard definition of the signal-to-noise ratio is based on the equality (10)

$$r_{S/N} = \sigma^2 / \sigma^2_{noise} \tag{S.1}$$

where $\sigma^2$ corresponds to the variance of primary variables without noise, whereas $\sigma^2_{noise}$ corresponds to the noise intensity. The variance $\sigma^2$ remains nearly constant at the increase of the number of repeating patterns, while the heights of related peaks will rapidly grow (cf., e.g., DFT spectra in Fig. 1D and in Fig. 2). If the equality (13) is fulfilled, the signal-to-noise ratio can be redefined using the mean intensity over equidistant peaks in DFT spectrum (Eq. (18))

$$\tilde{r}_{S/N} = N_P \sigma_P^2 / \sigma^2_{noise} \tag{S.2}$$

Such definition takes into account the growth of peaks with the increase of repeat numbers, though the applicability of definition (S.1) is restricted by the equality (13). Below, we will use the both definitions depending on the problem concerned. The statistically significant signal detection depends not only on signal-to-noise ratio but also on the length of data set $N$ and the shape of repeating patterns. If the number of periods is fixed and the features of repeats are re-scaled (such that $\sigma^2$ is approximately retained), the longer the data set, the lower the detection threshold for the signal-to-noise ratio.

*S.1. Large-scale noisy patterns with integer number of periods*

The examples in simulations were selected for the sake of clear presentation and cross-check. We begin with the simplest situation when the data set comprises the integer number of large-scale periods. Fig. S1 illustrates DFT and DDFT spectra for six periods of $10^4$ points. The periodic pattern corresponds to a sequence of pulses modeled by Boolean variables taking the values 0 and 1 randomly distributed within the period. The fraction of units was 0.1. The assessment of periods via the highest harmonic in DFT spectrum would be wrong for this particular example even in the absence of noise, whereas the peaks in DDFT spectrum provide the correct result. The number of periods assessed via the extreme left peak in equidistant series in DDFT spectrum was calculated by Eq. (25). The associated equidistant series for the related generating harmonic is also clearly seen in DDFT spectra (shown by arrows in Fig. S1) and provides an additional cross-check for the detection of noisy repeating patterns. For the signal-to-noise ratio $r_{S/N} = 0.033$ the standard deviation for noise exceeds that for signal 5.5 times. Nevertheless, even such strong noise does not suppress the related peaks in DDFT spectrum.

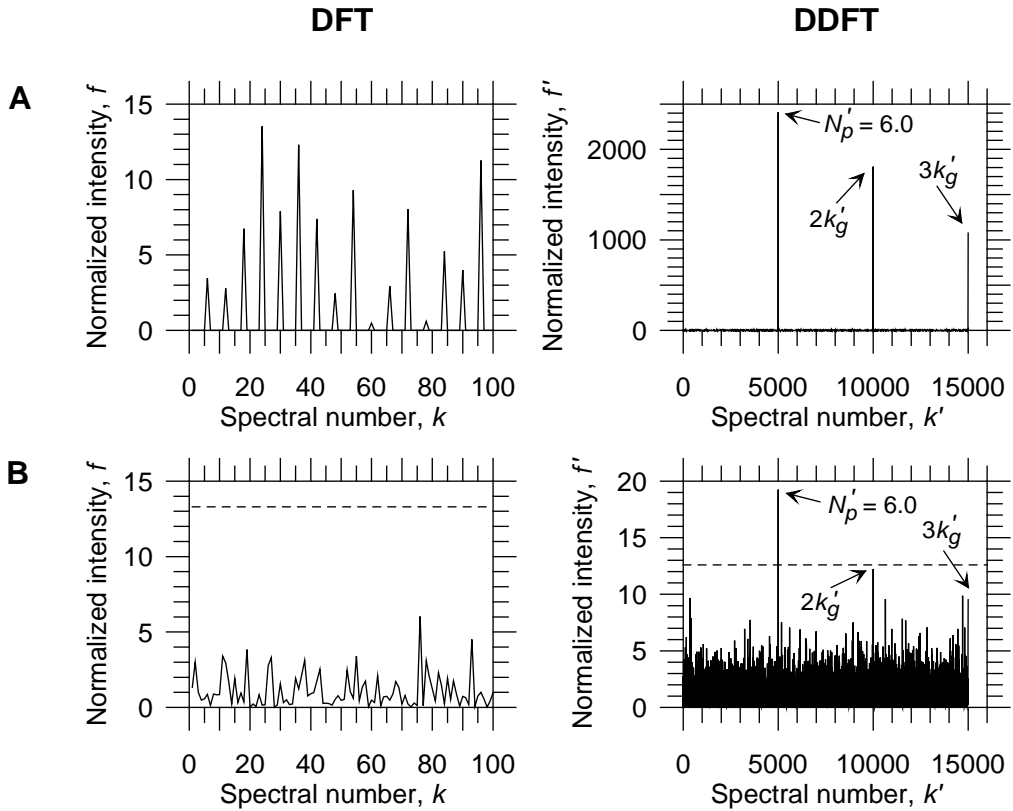

Fig. S1. The initial ranges of DFT spectra (left) and DDFT spectra (right) for the repeating sequence of pulses modeled by Boolean variables taking the values 0 and 1 randomly distributed within the interval of length $10^4$. The fraction of units was 0.1. The each periodic pattern was repeated 6 times. A, spectra for the patterns without noise; and B, spectra for the patterns with a particular realization of noise. The signal-to-noise ratio is slightly above the detection threshold, $r_{S/N} = 0.033$ (Eq. (S.1)) or $\tilde{r}_{S/N} = 0.2$ (Eq. (S.2)). The dashed horizontal lines in the panel B mark the extreme value thresholds corresponding to the probability Pr = 0.05 for Rayleigh spectra. The number of periods detected by the peak in the DDFT spectrum is given by Eq. (25). The associated equidistant peaks in DDFT spectrum are also shown (Eq. (24)).

*S.2. Fractional large-scale noisy patterns and their superposition*

Remarkably, DDFT spectra allow to resolve the fractional number of large-scale noisy repeats as well. We are not aware of any other method capable of such direct detection. We took the same periods of $10^4$ points as in Section S.1. In the first example the two periods of $10^4$ points were supplemented by 0.7 fragment of the period, whereas in the second example the six such periods were supplemented by 0.4 fragment of the period. The related DFT and DDFT spectra are shown in Figs. S2 and S3.

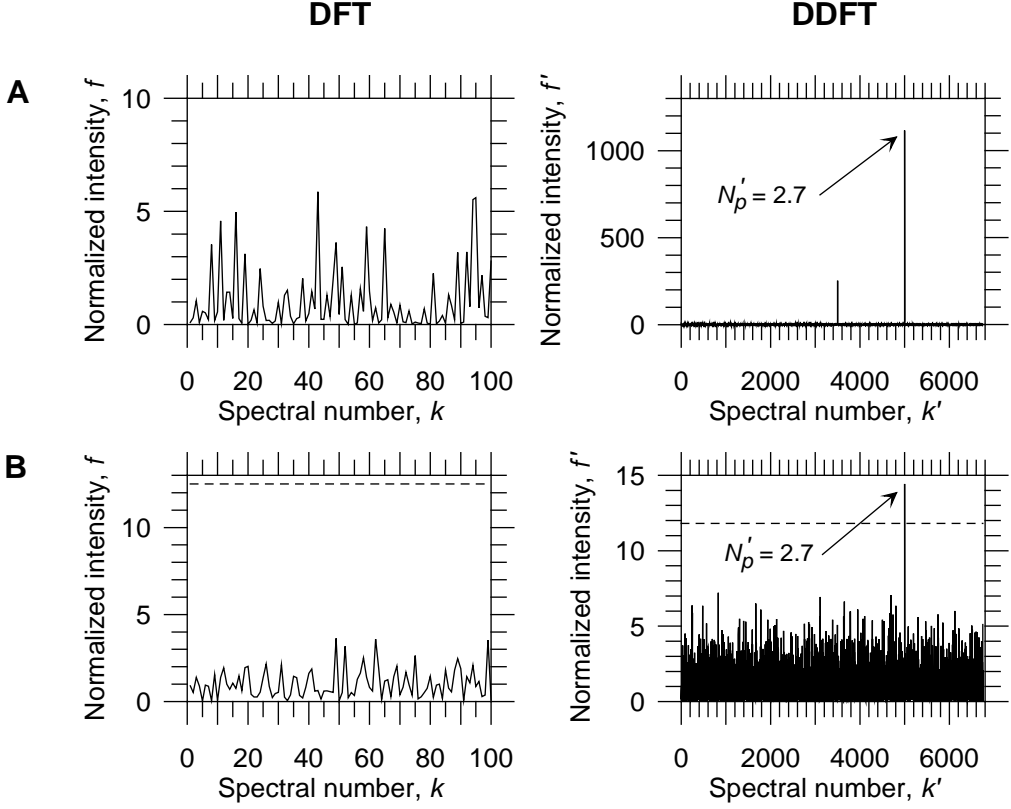

Fig. S2. The initial ranges of DFT spectra (left) and DDFT spectra (right) for the repeating sequence of pulses modeled by Boolean variables taking the values 0 and 1 randomly distributed within the interval of length $10^4$. The pattern was composed of two complete repeats and supplemented by 0.7 fragment of repeat. A, spectra for the patterns without noise; and B, spectra for the patterns with a particular realization of noise. The signal-to-noise ratio is slightly above the detection threshold, $r_{S/N} = 0.05$ (Eq. (S.1)). The dashed horizontal lines in the panel B mark the extreme value thresholds corresponding to the probability Pr = 0.05 for Rayleigh spectra. The number of periods detected by the peak in DDFT spectrum is given by Eq. (25).

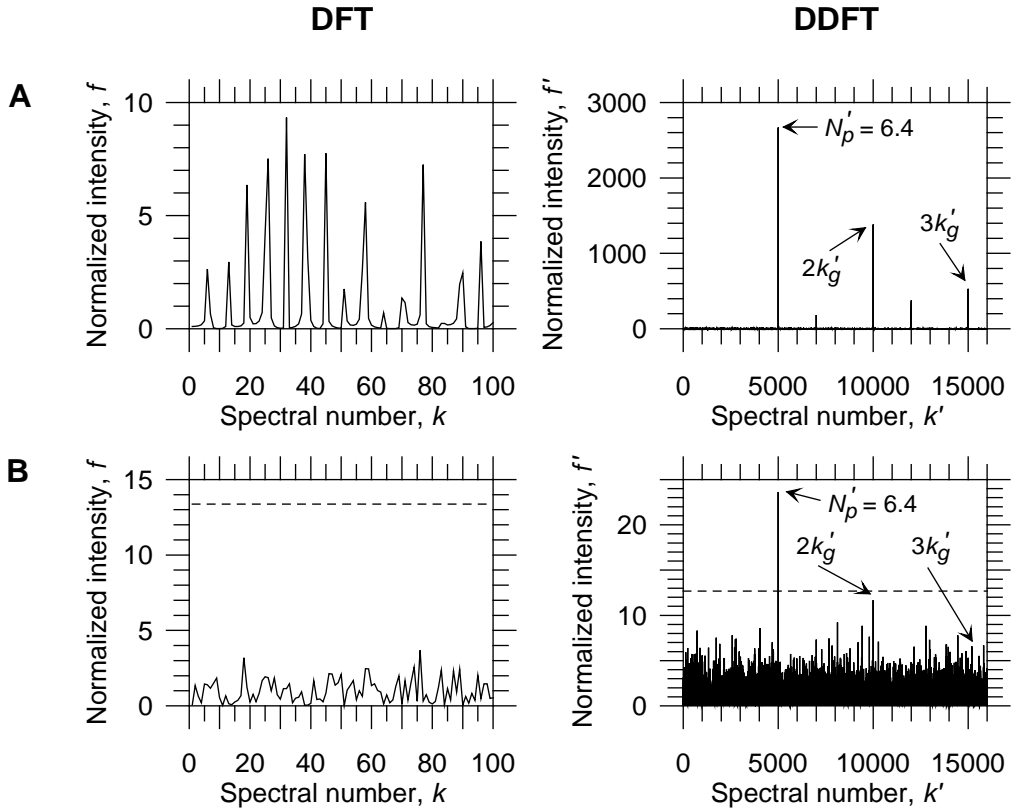

Fig. S3. The initial ranges of DFT spectra (left) and DDFT spectra (right) for the repeating sequence of pulses modeled by Boolean variables taking the values 0 and 1 randomly distributed within the interval of length $10^4$. The pattern was composed of six complete repeats and supplemented by 0.4 fragment of repeat. A, spectra for the patterns without noise; and B, spectra for the patterns with a particular realization of noise. The signal-to-noise ratio is slightly above the detection threshold, $r_{S/N} = 0.04$ (Eq. (S.1)). The dashed horizontal lines in the panel B mark the extreme value thresholds corresponding to the probability Pr = 0.05 for Rayleigh spectra. The number of periods detected by the peak in DDFT spectrum is given by Eq. (25). The associated equidistant peaks in DDFT spectrum are also shown (Eq. (24)).

The detection thresholds for the signal-to-noise ratio were close in both cases and were higher than those in the case of patterns with the integer number of periods. Then, we took the superposition of these fractional patterns, $x'_n = x_n^{(p_1)} + 0.9 x_n^{(p_2)}$ (with $N_{p1} = 2.7$ and $N_{p2} = 6.4$), on the total interval of $6 \times 10^4$ points. The DDFT spectra shown in Fig. S4 reveal that DDFT resolves the number of periods for the superposition as well. The corresponding signal detection threshold turns out to be higher for the superposition in comparison with the thresholds for the constituent fractional patterns. The precise determination of the number of patterns is important for the correct assessment of periods and the interpretation of underlying data.

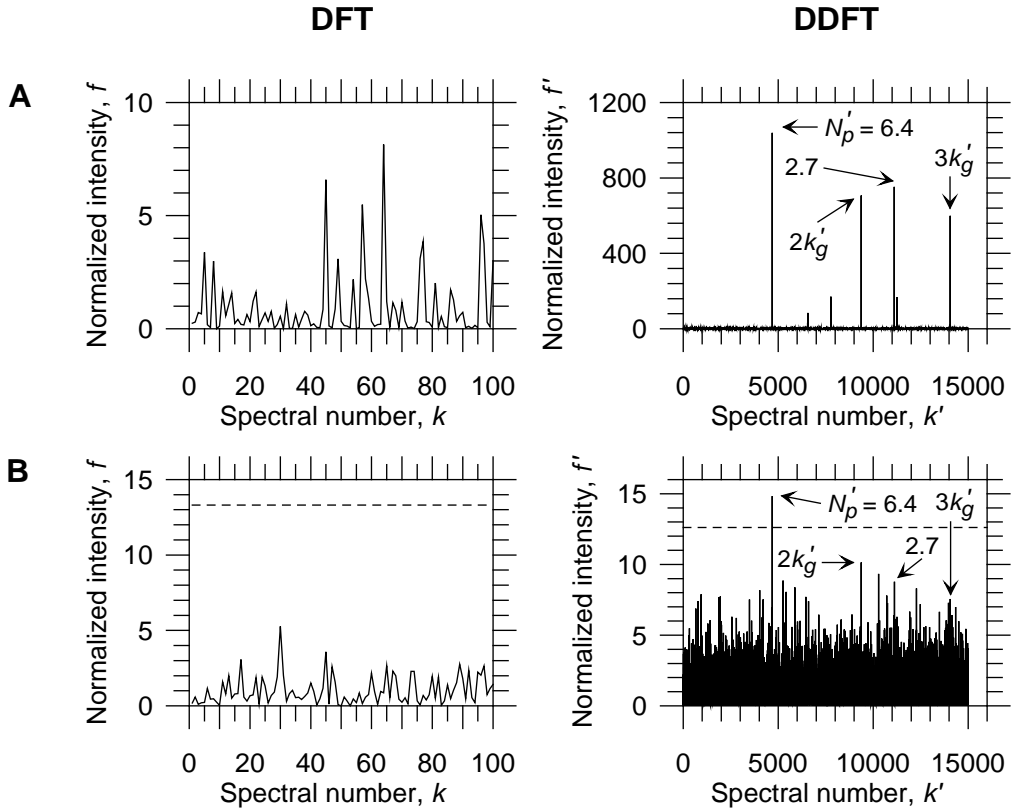

Fig. S4. The initial ranges of DFT spectra (left) and DDFT spectra (right) for the superposition of pulses modeled by Boolean variables taking the values 0 and 1 distributed within the set of $6\times10^4$ points. The superposition of pulses was chosen as $\Delta x'_n = \Delta x_n^{(p_1)} + 0.9\Delta x_n^{(p_2)}$ with the number of the first periods $N_{p1} = 2.7$ and the number of the second periods $N_{p2} = 6.4$. A, spectra for the patterns without noise; and B, spectra for the patterns with a particular realization of noise. The signal-to-noise ratio is slightly above the detection threshold, $r_{S/N} = 0.08$ (Eq. (S.1)). The dashed horizontal lines in the panel B mark the extreme value thresholds corresponding to the probability Pr = 0.05 for Rayleigh spectra. The number of periods detected by the peaks in DDFT spectrum is given by Eq. (25). The associated equidistant peaks for the period $p_2$ in DDFT spectrum are also shown (Eq. (24)).

*S.3. Superposition of short and large-scale noisy periodic patterns*

In this section we will discuss peculiarities inherent to search for large-scale noisy periodic patterns in superposition with short periods. The relevant preprocessing is illustrated with the superposition $x'_n = x_n^{(p_1)} + 0.3 x_n^{(p_2)}$ (with $N_{p1} = 6\times10^3$ and $N_{p2} = 6$) on the total interval of $6\times10^4$ points. The related DFT and DDFT spectra without noise and preprocessing are shown in Fig. S5A. In this case the peaks in DDFT spectrum corresponding to the large-scale periodicities are barely discernible on the background of frequent peaks for the short periodicities. The broad trend in the background reflects the uncertainty principle for Fourier transform. This broad trend in DDFT spectrum is generated by the rare high peaks in DFT spectrum related to the short periodicities.

To search for the large-scale periodicities in superposition with the short periods, DFT spectra should be preliminarily preprocessed. We used the cut-off of the highest peaks by the characteristic outbursts in Rayleigh spectra assessed by extreme value statistics, $\ln(K/0.05)$, where $K$ is defined by Eq. (9). If the height of a harmonic in DFT spectrum exceeds the threshold $\ln(K/0.05)$, the height is replaced by the threshold. Then, DDFT is performed for such preprocessed DFT spectrum. After preprocessing, the peaks for the large-scale periodicities in DDFT spectrum become clearly resolved (Fig. S5B). The same preprocessing was applied to the noisy patterns (Fig. S5C). Although the resulting signal detection threshold for the superposition was significantly higher than the threshold determined in Section S.1 for the pure large-scale patterns (Fig. 4B), the formal replacement of the total variance corresponding to the superposed signal by the contribution related to the large-scale periodicities would provide the threshold not far from that in Section S.1.

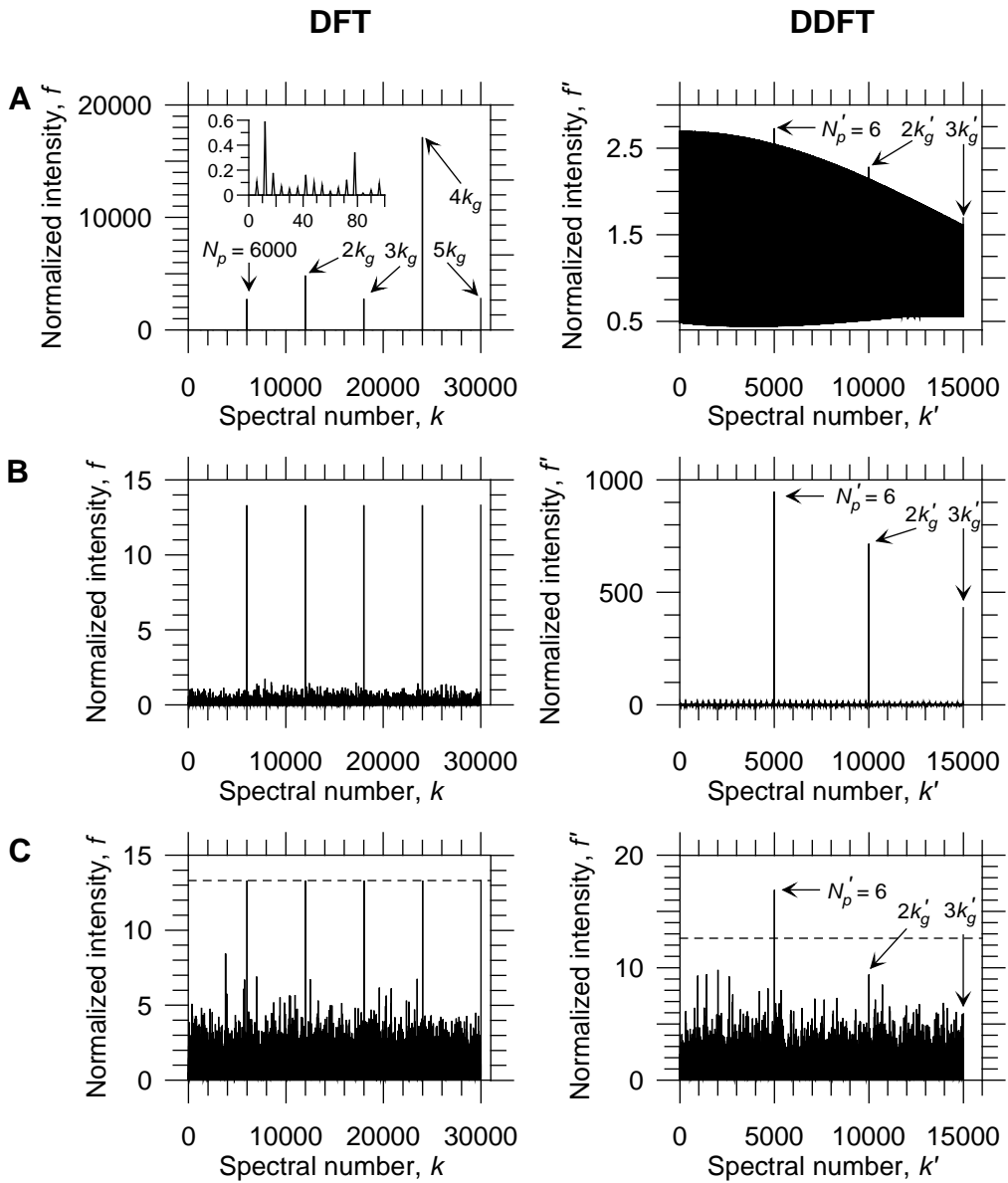

Fig. S5. DFT and DDFT spectra for the superposition of pulses modeled by Boolean variables taking the values 0 and 1 distributed within the set of 6×10⁴ points. The superposition of pulses was chosen as $\Delta x'_n = \Delta x_n^{(p_1)} + 0.3 \Delta x_n^{(p_2)}$ with the number of the first periods $N_{p1} = 6\times10^3$ and the number of the second periods $N_{p2} = 6.0$. A, spectra for the patterns without noise; B, spectra for the patterns without noise; peaks in DFT spectrum were preliminary cut-off as described in the main text; then, DDFT was performed for DFT spectrum with cut-off peaks; and C, spectra for patterns with a particular realization of noise. The signal-to-noise ratio is slightly above the detection threshold, $r_{S/N} = 0.5$ (Eq. (S.1)). In the panel C peaks in DFT spectrum were preliminary cut-off as described in the main text; then, DDFT was performed for DFT spectrum with cut-off peaks. The dashed horizontal lines in the panel C mark the extreme value threshold corresponding to the probability Pr = 0.05 for Rayleigh spectra. The number of periods detected by the peaks in DDFT spectrum is given by Eq. (25). The equidistant series for the period $p_1$ in DFT spectrum is shown in the panel A (left), whereas the equidistant peaks for the period $p_2$ in DDFT spectra are shown on the right (Eq. (24)).

To simulate the effects of true modulations, the periodic Gaussian noise with periods $p = 10^3$ was added directly to the unit pulses related to the short periods $p = 10$. The standard deviation for the periodic Gaussian noise was 0.1. The cut-off procedure and the subsequent addition of noise were as described above. We did not observe any differences with the spectra presented in Fig. S5. The threshold signal-to-noise ratio was in this case $r_{S/N} = 0.2$.

*S.4. Method of periodograms*

The sensitivity of signal detection in the presence of noise can be enhanced by averaging over $N_s$ counterpart spectra (method of periodograms [12]). At the limit $r_{S/N} \ll 1$, the noise dominates over signal and the normalized deviations (48) are applicable to the signal detection. At this limit the following asymptotic relationship can be derived for the normalized deviations

$$\varsigma_k \approx r_{S/N}(< f_k >_{signal} -1)/(1/N_s)^{1/2} \qquad (S.3)$$

where the angular brackets denote averaging over spectra and $< f_{k,signal} > = < F_{k,signal} >/\sigma^2_{signal}$. We wrote $< F_{k,signal} >$ because the interference between noise and signal during averaging over finite number of spectra is not completely suppressed. As the normalized DDFT harmonics (22) remain invariant under affine transforms of DFT intensity harmonics (6), Eq. (S.3) indicates that DDFT spectrum for signals can be completely restored at the limit $N_s \to \infty$ even for small signal-to-noise ratios, $r_{S/N} \ll 1$.

We compared two processing modes: (i) DFT → averaging → DDFT is preferable over (ii) DFT → DDFT → averaging, i.e. in (i), first, DFT is calculated, then, DFT spectra are averaged over random realizations and, finally, DDFT spectrum is calculated for the resulting averaged DFT spectrum, whereas in (ii), first, DFT is calculated, then, DDFT and, finally, DDFT spectra are averaged over random realizations. We chose the same superposition of fractional large-scale patterns as in Section S.2. The relevant spectra without noise and averaging are shown in Fig. S4A. The results of simulations performed to assess the signal detection thresholds for the two processing modes are presented in Fig. S6. The shown spectra correspond to the values of signal-to-noise ratio $r_{S/N}$ slightly above detection threshold. The DDFT spectra in Fig. S6 reveal clear preference of processing mode (i) over (ii). For the both modes the dependence of detection threshold on the number of averaged spectra, $N_s$, is in approximate agreement with theoretical prediction, $r_{S/N} \propto N_s^{-1/2}$. We compared also the dependence of detection threshold for averaging of non-normalized and normalized DFT spectra in the mode (i) and did not find any meaningful difference. Nevertheless, in the presence of additional large-scale variations within a set of averaged spectra the mode with normalized DFT spectra seems to be preferable. For symbolic sequences such normalization would suppress the dependence of periodicities on the variations in the symbol content within processed sequences [8, 23]. For the lowest signal-to-noise ratio, $r_{S/N} = 0.0008$ (Fig. S6D), the standard deviation for the noise would be 35.4 times higher than that for the signal.

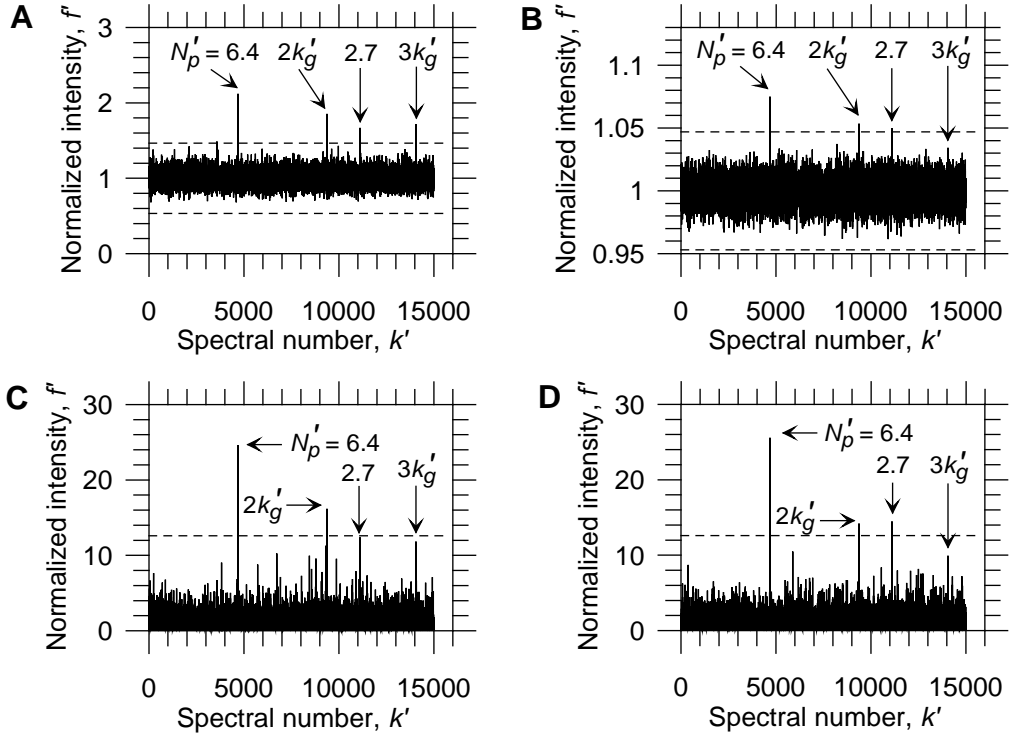

Fig. S6. DDFT spectra for the different modes of averaging. The panels A and B correspond to the processing mode: DFT → DDFT → averaging, whereas the panels C and D correspond to the mode: DFT → averaging → DDFT. The superposition of periodic patterns was the same as that in Section S.2 (see also Fig. S4A). The particular parameters for the spectra shown are: A, $N_s = 100$, $r_{S/N} = 0.02$; B, $N_s = 10000$, $r_{S/N} = 0.005$; C, $N_s = 100$, $r_{S/N} = 0.008$; and D, $N_s = 10000$, $r_{S/N} = 0.0008$. The dashed horizontal lines in the panels A and B mark the extreme value thresholds corresponding to the probability Pr = 0.05 for the Gaussian deviations (48), while the dashed horizontal lines in the panels C and D mark the extreme value thresholds corresponding to the probability Pr = 0.05 for Rayleigh spectra.

*S.5. Large-scale periodic patterns in the presence of noise with long-range correlations*

White noise considered above is not a unique source of interference in the experimental spectra. In this section we explain why DDFT is especially important for detection of large-scale quasi-periodic patterns in the presence of noise with long-range correlations. The correlations in noise were generated as follows. First, the Gaussian noise was generated at the each sampling point $n$. Then, the noise was consecutively smoothed within the window of width $w$ sliding with step one along the data set. Such smoothing produces the correlation radius $R_{corr} \approx w$ in the noise. The details of statistical theory for DFT of the smoothed Gaussian noise are presented in Appendix C. DFT for the correlated data yields the high harmonics in the range of low spectral numbers with the heights diminishing with the increase of spectral number [8]. The characteristic range for such a trend in DFT spectrum is $k_{trend} \approx N/R_{corr} \approx N/w$. Furthermore, such trend in DFT spectrum will generate a related trend in DDFT spectrum according to uncertainty principle for the discrete Fourier transform. The range for the related trend in DDFT spectrum can be estimated as $k'_{II,trend} \approx K/k_{trend} \approx w/2$. The high harmonics produced by the correlated noise in the range of low spectral numbers in DFT spectrum would be superimposed on the counterpart harmonics produced by the signal and might incorrectly be attributed to large-scale quasi-periodic patterns. DDFT is able to distinguish among contributions from correlated noise and the periodic patterns in DFT spectra. It is also important that the high peaks generated in DDFT spectrum by the large-scale quasi-periodic patterns are commonly beyond the trend range induced by the correlated noise.

The clusters of high harmonics in separate parts of normalized Fourier spectrum diminish the heights of harmonics in other parts of the spectrum and lead to the underestimation of the statistical significance of periodicities. Therefore, the range of DDFT spectrum beyond the trend range should be renormalized to improve the statistical assessment. Let the spectral range corresponding to the normalized harmonics (22) with the spectral numbers $k'_{II} \in (k'_1, k'_2)$ be approximately homogeneous. Calculate the local mean value within this range,

$$\bar{f}_{k'_1 k'_2; II} = \frac{1}{k'_2 - k'_1 + 1} \sum_{k'=k'_1}^{k'_2} f_{k'; II} \tag{S.4}$$

and renormalize the harmonics (22) according to

$$f^{(renorm)}_{k'; II} = f_{k'; II} / \bar{f}_{k'_1 k'_2; II}; \ k' \in (k'_1, k'_2) \tag{S.5}$$

DFT and DDFT spectra for the Gaussian noise smoothed within the sliding window of width $w = 1.1 \times 10^3$ on the data set of $6 \times 10^4$ points are shown in Fig. S7A. The characteristic breadth of clustering in DDFT spectrum is about $\Delta k'_{II} \approx w/2$ as estimated above. The patterns composed of six periods were the same as those in Section S.1 (see also Fig. S1). Then, the correlated noise was added to these patterns as described above. The corresponding DFT and DDFT spectra near detection threshold are shown in Fig. S7B. The threshold signal-to-noise ratio for the correlated noise turns out to be nearly an order of magnitude higher than that for the white noise.

The threshold can, however, be drastically reduced by the preprocessing of DFT spectra. It is worth, first, de-trending DFT harmonics as

$$f_k^{(de-trend)} = f_k / \text{reg} < f_k^{(w)} > \tag{S.6}$$

where

$$\text{reg} < f_k^{(w)} > = \begin{cases} < f_k^{(w)} >, \text{if} < f_k^{(w)} > \geq 0.01 \\ 0.01, \text{if} < f_k^{(w)} > < 0.01 \end{cases} \tag{S.7}$$

and the harmonics $< f_k^{(w)} >$ are defined by Eq. (52). Then, DDFT should be applied to harmonics (S.6). After de-trending, the renormalized ranges of DDFT spectra reveal the peaks related to periodic patterns at the much lower signal-to-noise ratios (Fig. S7C). Additionally, DDFT displays also the peak related to the rapidly diminishing variations in $< f_k^{(w)} >$ at $k'_{II} = (K-1)/(N/w) \approx w/2$ (see Fig. S7C and Eq. (52)). The relevant periodograms are shown in Fig. S8. For the lowest signal-to-noise ratio, $r_{S/N} = 2 \times 10^{-5}$, the standard deviation for the noise would be 223.6 times higher than that for the signal.

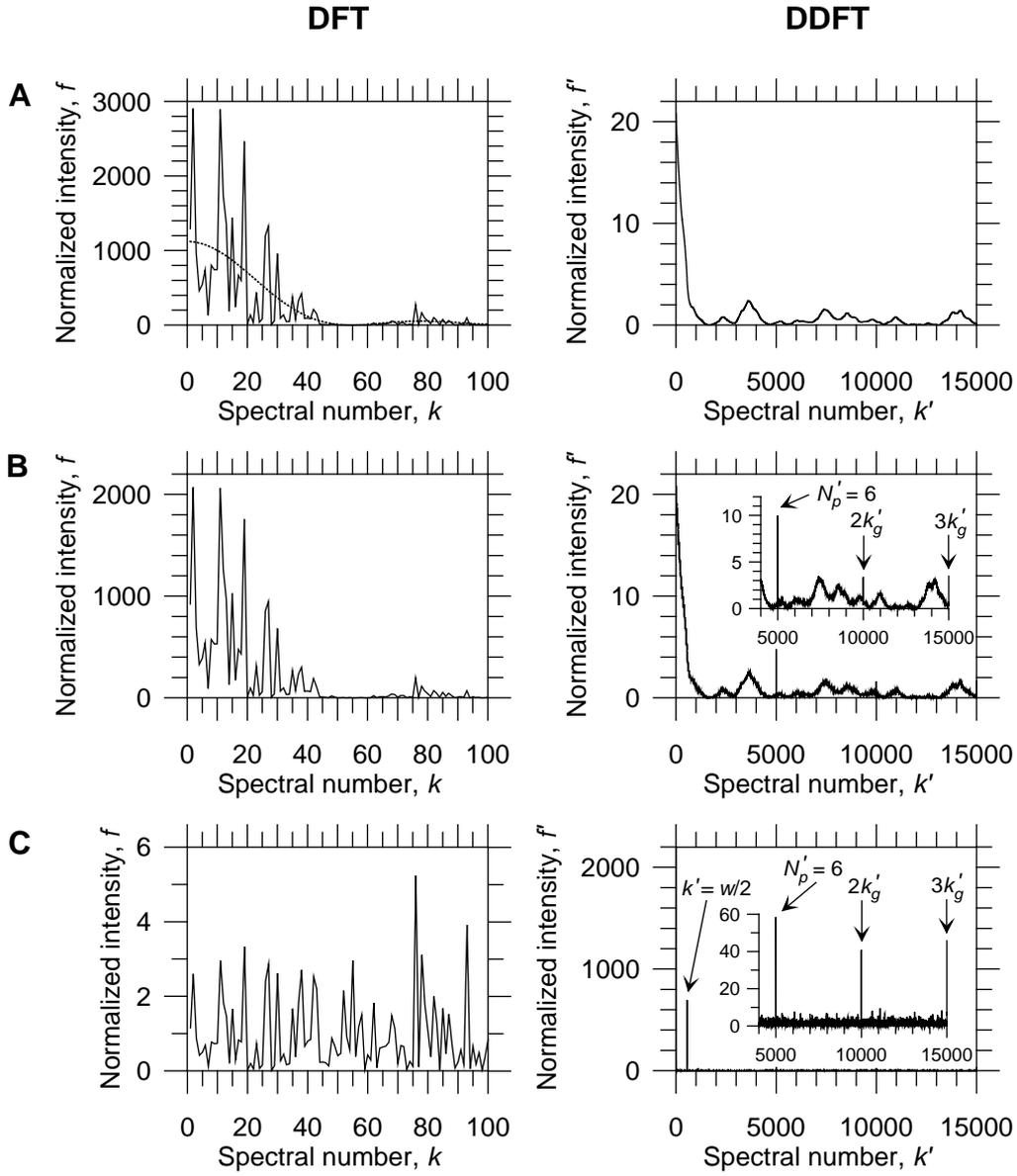

Fig. S7. The effect of noise with long-range correlations on the detection of large-scale periodic patterns. The periodic patterns were the same as those in Section S.1 (see also Fig. S1). The correlated noise was generated by the smoothing of white noise within the sliding window of width $w = 1.1 \times 10^3$. A, the initial range of DFT spectrum (left) and DDFT spectrum (right) for a particular realization of correlated noise. The dotted line in DFT range shows the mean intensities over many realizations (Eq. (52)). B, the initial range of DFT spectrum (left) and DDFT spectrum (right) for the periodic patterns with a particular realization of correlated noise. The signal-to-noise ratio is slightly above the detection threshold, $r_{S/N} = 0.4$ (Eq. (S.1)) or $\tilde{r}_{S/N} = 2.4$ (Eq. (S.2)). The insert in the DDFT spectrum corresponds to the renormalized range (Eq. (S.5)). C, the initial range of DFT spectrum (left) and DDFT spectrum (right) after de-trending of DFT spectrum (Eq. (S.6)) for the periodic patterns with a particular realization of correlated noise. The signal-to-noise ratio is slightly above the detection threshold, $r_{S/N} = 0.0002$ (Eq. (S.1)) or $\tilde{r}_{S/N} = 0.0012$ (Eq. (S.2)). The insert in the DDFT spectrum corresponds to the renormalized range (Eq. (S.5)).

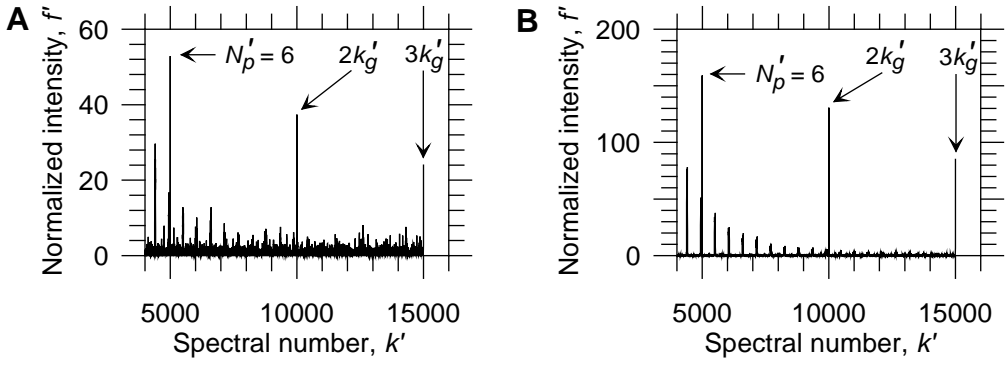

Fig. S8. The periodograms for the periodic patterns in the presence of noise with long-range correlations. The periodic patterns were the same as those in Section S.1 (see also Fig. S1). The noise was generated as described in Section S.5. The periodograms were obtained by the following processing scheme: normalized DFT for the periodic patterns with a particular noise realization → averaging over noise realizations → de-trending of averaged DFT spectrum by Eq. (S.6) → DDFT → choosing the spectral range 4000–end and its renormalization by Eq. (S.5). The signal-to-noise ratio is $r_{S/N} = 2 \times 10^{-5}$ (Eq. (S.1)) or $\tilde{r}_{S/N} = 1.2 \times 10^{-4}$ (Eq. (S.2)). A, 100 noise realizations. B, 10000 noise realizations.

Although the effect of correlated noise on the signal detection without additional preprocessing was stronger in comparison with that for the white noise, the situation reciprocates after preprocessing of DFT spectra. The white noise acts approximately uniformly over all spectral ranges, whereas the impact of noise with long-range correlations is mainly squeezed in a range of the low spectral numbers in DFT spectra. Therefore, the suppression of correlated noise by de-trending harmonics in the range of the low spectral numbers in DFT spectrum strongly improves the detection of large-scale quasi-periodic patterns by DDFT. Though the choice of the particular de-trending in our example depended on the chosen model for correlated noise, the conclusion on the obligatory de-trending in the range of the low spectral numbers in DFT spectrum is general for the improvement of the search for noisy quasi-periodic patterns via DDFT. The smoothing of initial data $\{x_n\}$ is often used for detection of large-scale periods. In the presence of noise such smoothing may induce high harmonics in DFT spectrum related to noise rather than to periodic patterns and actually may hamper detection of the patterns.